\documentclass[journal,keeplastbox]{IEEEtran}
\usepackage[T1]{fontenc}
\usepackage[utf8]{inputenc}
\usepackage{color}
\usepackage{float}
\usepackage{braket}
\usepackage{mathrsfs}
\usepackage{amsmath}
\usepackage{amsthm}
\usepackage{amssymb}
\usepackage{graphicx}
\usepackage[export]{adjustbox}
\theoremstyle{plain}
\usepackage{amsmath}
\usepackage{amsfonts}
\usepackage{bm}
\usepackage{setspace}
\usepackage{cite}
\usepackage{subcaption}
\usepackage{threeparttable}
\usepackage[table]{xcolor}
\usepackage{multirow}
\usepackage{algorithm}
\usepackage{algpseudocode}

\newcommand{\vectnorm}[1]{\left\lVert\begin{bmatrix}#1\end{bmatrix}\right\rVert_2}
\newcommand{\blue}[1]{{\textcolor[rgb]{0,0,0}{#1}}}
\algnewcommand\algorithmicforeach{\textbf{for each}}
\algdef{S}[FOR]{ForEach}[1]{\algorithmicforeach\ #1\ \algorithmicdo}

\allowdisplaybreaks
\title{Multi-IRS Aided Mobile Edge Computing for High Reliability and Low Latency Services}

\author{Elie El Haber, Mohamed Elhattab, Chadi Assi, Sanaa Sharafeddine, and Kim Khoa Nguyen}

\begin{document}
	
 \maketitle
 
{\let\thefootnote\relax\footnotetext{A preliminary part of this work has been published in IEEE International Conference on Communications 2022 \cite{Elie_ICC}. Specifically, in \cite{Elie_ICC}, we study the optimized use of the IRS along with the design of the offloading and resource allocation parameters for maximizing the UEs’ sum of offloaded bit. 
}}

\begin{abstract}
    Although multi-access edge computing (MEC) has allowed for computation offloading at the network edge, weak wireless signals in the radio access network caused by obstacles and high network load are still preventing efficient edge computation offloading, especially for user requests with stringent latency and reliability requirements. Intelligent reflective surfaces (IRS) have recently emerged as a technology capable of enhancing the quality of the signals in the radio access network, where passive reflecting elements can be tuned to improve the uplink or downlink signals. Harnessing the IRS's potential in enhancing the performance of edge computation offloading, in this paper, we study the optimized use of a system of multi-IRS along with the design of the offloading (to an edge with multi MECs) and resource allocation parameters for the purpose of minimizing the devices' energy consumption considering 5G services with stringent latency and reliability requirements. After presenting our non-convex mathematical problem, we propose a suboptimal solution based on alternating optimization where we divide the problem into sub-problems which are then solved separately. Specifically, the offloading decision is solved through a matching game algorithm, and then the IRS phase shifts and resource allocation optimizations are solved in an alternating fashion using the Difference of Convex approach. The obtained results demonstrate the gains both in energy and network resources and highlight the IRS's influence on the design of the MEC parameters.
\end{abstract}


\begin{IEEEkeywords}
Computation offloading, intelligent reflecting surface, multi-access edge computing, ultra-reliable low-latency communication.
\end{IEEEkeywords}

\section{Introduction}\label{sec:intro}
The advent of fifth-generation and beyond (B5G) networks has been accompanied by a widespread proliferation of data and computation-hungry services that require the computation of the collected input and have extreme requirements for quality metrics such as latency, reliability, and energy. A subset of those services are the ones belonging to the ultra-reliable and low-latency communication (URLLC) category, which have imposed tremendous pressure on enhancing existing network infrastructures for those services to be provisioned seamlessly in B5G networks~\cite{siddiqi20195g}. Towards that end, multi-access edge computing (MEC) came into the picture as a novel technology that would assist in lowering the computation latency and increasing the reliability of data computation, in addition to reducing the energy load on mobile devices that would normally perform the computation procedure. This is done by offloading the computation to servers (called cloudlets) that are co-located within the base stations (BSs) on the network edge~\cite{mao2017survey,hu2015mobile}.

Although MEC has proved thus far to be a worthy technology for enabling modern B5G services with their stringent requirements, there are limitations concerning the radio access network (RAN) communication quality that are still hindering the full realization of MEC as a sustainable solution in next-generation wireless networks. Those limitations are the product of a poor communication environment caused by blockages, situations of peak load, or under-served areas, which cause the access channel quality to be degraded. This situation causes a high delay to be incurred when trying to conduct the offloading operation, which either forces the limited devices to spend high offloading energy to satisfy the service requirements or prevents the offloading procedure from happening altogether while wasting the utilization of edge resources and keeping them under-utilized. The problem is more exaggerated when trying to satisfy stringent requirements, such as those of services requiring high-reliability and low-latency 
\cite{siddiqi20195g}.

Multiple technologies have been explored so far in conjunction with MEC for enhancing the offloading performance, such as the utilization of unmanned aerial vehicles (UAV), small cells, and non-orthogonal multiple access (NOMA), where each has its advantages and limitations when used in the context of edge computing~\cite{yazid2021uav}. Owing to recent progress in programmable meta-materials, a recent introduction into mobile networks is intelligent reflecting surfaces (IRSs) which can passively enhance the RAN wireless communications quality through exploiting a high number of low-energy low-cost reflecting elements~\cite{wu2021intelligent}. With the necessary optimizations, the amplitude and/or the phase shift of the IRS's elements can be tuned in order to enhance the wireless propagation environment, as has been demonstrated in~\cite{wu2019towards,wu2019intelligent}.

Now, due to the observed difficulty in enabling energy-efficient MEC computation offloading for services with low latency and high reliability in a network with unfavorable wireless channels, we envision the introduction of the IRS to play a significant role in reducing the overall devices' energy consumption while satisfying their strict requirements. \blue{Particularly, since the IRS can improve the channel conditions and hence the transmission rates and upload latency, less offloading energy can be spent by the user equipments (UEs). In addition, since this reduction in upload latency can give more flexibility for having a higher computation latency, a reduction in MEC resource utilization can be obtained, which will give room for other tasks to use those saved resources and hence accommodate more UEs. Thus, an IRS-aided MEC system would enable devices to access MEC servers (that could offer a higher amount of resources) which would normally require a high transmission power or even be unreachable due to poor-quality channels.} However, in scenarios where the UEs need to offload to multiple cloudlets due to their reliability requirements, the judicious sharing of the IRS among the offloading users and the phase shifts optimization while minimizing the overall UEs' energy consumption is a challenging problem that we aim to tackle.

In this paper, we investigate the problem of a multi-user MEC computation offloading in an orthogonal frequency division multiple access (OFDMA) system where the IRS is utilized for the purpose of minimizing the overall energy consumption while satisfying the services' low latency and high-reliability requirements. We consider MEC servers' reliability and task redundancy, where the task of each UE can be simultaneously offloaded to multiple servers to guarantee its computational reliability. In this context, we are interested in studying the impact that the IRS optimization has on the design of the other MEC parameters. Our contributions can be summarized as follows:
\begin{enumerate}
    \item We seek to jointly optimize the IRS elements' phase shift, the offloading decision, the UE's transmit power, as well as the allocated OFDMA resource block and server computational resources; in addition to guaranteeing the service quality required by the computing task (i.e., latency and reliability), we aim at minimizing the total offloading energy consumption. We mathematically model this problem and formulate it as a non-convex Mixed-Integer Nonlinear Program (MINLP).
    \item We propose a customized suboptimal solution based on the alternating optimization approach, where the problem is divided into multiple sub-problems. First, the offloading decision is performed using a matching game algorithm. Then, the other parameters' design is performed through two sub-problems that are solved iteratively. On one hand, the IRS elements' phase shifts are optimized through a novel technique based on the successive convex approximation (SCA) approach, which provides an approximate solution by iteratively solving until convergence. This is done after the problem is converted to a difference of convex (DC) representation~\cite{marks1978general,10109654, 9586734}, and the results are then used as inputs for the second sub-problem. On the other hand, the UEs' transmit power and resource allocation parameters are also optimized through an SCA-based approach after the problem is converted to a Second Order Cone Program (SOCP). The two sub-problems are solved in an alternating fashion until the objective convergence.
    \item We present numerical results considering various system configurations where the solution performance is demonstrated; elaborate studies are presented to showcase the impact of optimizing the IRS elements in reducing the UE energy consumption while satisfying their service quality.
\end{enumerate}

The use, together with their optimal configuration, of IRS elements would allow devices to effectively access edge resources, which may be hard to reach due to poor channel quality to the base stations hosting these computing resources, unless high transmit power is used. Our work provides insights for leveraging IRS-aided APs to reduce the energy consumption of UEs that are requesting services with low latency and high-reliability requirements such as mission-critical applications, and also for influencing the design of other MEC network parameters.

In the remainder of this paper, section \ref{sec:literature} presents the related work. Section \ref{sec:model} presents our system model and the proposed non-convex problem. Section \ref{sec:solution} presents the customized solution approach and the overall algorithm for solving the problem. Section \ref{sec:results} discusses the numerical analysis. Finally, section \ref{sec:conclusion} concludes the paper.

\section{Literature Review}\label{sec:literature}
We present first the main papers that explored the use of IRS for enhancing wireless communication in the access network. Then, we present studies that focus on the use of IRS in the context of edge computing for enhancing the computation offloading performance. Finally, we present related work that studied reliability and latency problems considering a MEC system. 

In \cite{wu2019intelligent}, the authors studied the use of one IRS in one cell where multiple single-antenna UEs are communicating with a multi-antenna AP, with the objective of minimizing the access point (AP) transmit power by optimizing the transmit beamforming and IRS phase shifts while respecting the users' required quality of service. In \cite{wu2019towards}, the authors provided an overview of the use of IRS in wireless networks and discussed its advantages, challenges, and architecture. Also, numerical results are presented that show the performance improvement brought by the use of the IRS.

In \cite{bai2020latency}, the authors proposed a block coordinate descent (BCD) based algorithm for minimizing the latency of computation offloading considering a set of users communicating with a multi-antenna AP, where the computation and communication resources are optimized, and the IRS's impact on the performance improvement is demonstrated. In \cite{liu2020intelligent}, the authors developed a solution for optimizing the IRS phase shifts for maximizing the operator's earnings in terms of the devices' payments while minimizing the users' weighted sum of latency and energy. In \cite{bai2021resource}, the authors developed a solution for performing wireless power transfer in an OFDMA system through the use of IRS in an edge computing setting where the operator's energy consumption is minimized. In \cite{chu2020intelligent}, the authors studied an IRS-aided MEC system where the IRS phase shift is optimized for maximizing the number of offloaded bits from a set of users to one AP. The authors in \cite{li2021energy} minimized the sum energy consumption through a BCD approach for a set of users communicating with a cloudlet-enabled AP where the IRS phase shift along with the offloaded data, transmit power, and time-division multiple access (TDMA) resources allocation are optimized considering non-orthogonal multiple access (NOMA). The authors in \cite{wu2021irs} maximized the computation offloading rate in an IRS-aided MEC system to explore the offloading performance considering the TDMA and NOMA multiple access schemes.

In \cite{liu2017latency}, the authors focused on minimizing the servers' transmit power for a set of users offloading with latency and reliability constraints.
\cite{liu2018offloading} studied the trade-off between latency and reliability in a MEC system, where the end-to-end latency and the failure probability of task offloading are minimized. More recent work \cite{xu2022energy} considered the offloading of tasks in an IRS-assisted NOMA network where the authors focused on minimizing the sum energy consumption by jointly optimizing the transmit powers, transmission time, offloading partitions and IRS phase shift. Their system model consisted of only one AP (and thus one IRS) and one MEC server, and they did not address the reliable and latency-aware offloading. The authors of \cite{zhou2022latency} assumed a VR content stored at the edge and to reduce latency for users, they considered an IRS-assisted network and security constraints. Their system model consisted of only one MBS and did not address reliable offloading. Similarly, other work (e.g., \cite{hu2021reconfigurable}, \cite{yang2022intelligent}, \cite{mao2022reconfigurable}, \cite{zhang2021drl}) leveraged IRS technology to assist MEC offloading but did not address the reliable offloading in a multi-AP, multi-IRS, and multi-MEC setting.

In our previous work~\cite{9563047}, we provided a first and rather simplistic attempt at studying the reliability-aware IRS-assisted computation offloading problem to guarantee the reliability and latency of the offloaded tasks, where one UE is considered in the context of one cloudlet-enabled AP. In contrast to the existing studies that considered IRS-aided MEC systems, in this work, we explore the use of multiple IRSs for helping to minimize the total energy consumption for a set of UEs while guaranteeing the tasks' stringent latency and reliability requirements. Here, redundancy is performed by simultaneously offloading the tasks to multiple servers to guarantee their computing reliability, where the optimized use of the IRSs which need to be judiciously shared in this scenario, are performed along with the offloading and resource allocation parameters. In this context, the IRS elements which influence the offloading latency can be optimized such as to reduce \blue{the MEC resources' usage,} and also to affect the design of the other MEC parameters for achieving a minimal energy consumption. Hence, the use of the IRSs is explored to cater to the limited devices' energy and capability, which would contribute towards enabling future mission-critical services in IoT networks.

\section{System Model}\label{sec:model}
\begin{figure}[!t]
    \centering
    \includegraphics[width=1\linewidth,trim={0.15cm 0.5cm 0.05cm 0.5cm},clip]{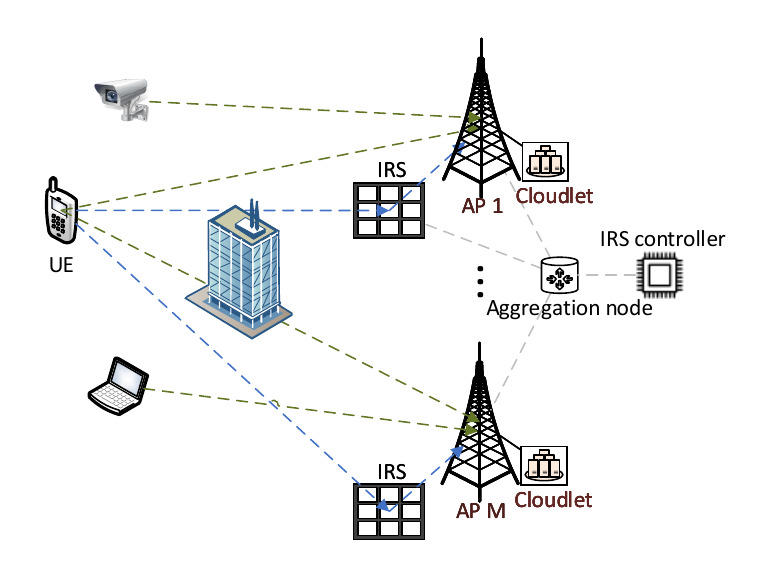}
    \caption{System model.}
    \label{fig:model2}
\end{figure}
    
\subsection{Network Model}
Our system model consists of a set of APs $\mathcal{M}$ ($M$ in total), serving a set $\mathcal{K}$ of devices or UEs ($K$ in total), as shown in Figure \ref{fig:model2}. Each device needs to offload a task to one AP (or more) where it receives a computing service from a co-located edge computing server or cloudlet. A cloudlet co-located with AP $j \in \mathcal{M}$ has a computational capacity $F_j$ (GHz) for serving UEs' requests. In addition, we assume each cloudlet $j\in\mathcal{M}$ to have an associated reliability, denoted by $\phi_j$.


A set of IRS surfaces, each comprised of $N$ reflecting elements, denoted by $\mathcal{N}_j=\{1,2,...,N\}$ is deployed for each AP $j$ in order to assist in offloading the tasks of UEs through enhancing the communication link between the UE and the associated AP \footnote{In this paper, we consider index $j$ for the IRS co-located with AP $j$.}. We define $x_{kj} \in \{0,1\}$, a decision variable indicating if UE $k$ is offloading its task to cloudlet $j\in\mathcal{M}$. The task of UE $k$ is defined by a tuple $\{D_k, C_k, \bar{L}_k, \bar{R}_k\}$, comprising the task size $D_k$ (Kbits), computing demand $C_k$ (CPU cycles per bit), required latency $\bar{L}_k$ (milliseconds), and required reliability $\bar{R}_k$.
    
\subsection{Communication Model}
We consider the uplink communication used by the UEs for task offloading and ignore the downlink communication since the task output size is in general much smaller than the task input size~\cite{wang2017joint}. We consider an OFDMA system with assumed perfect channel state information (CSI), where the radio spectrum is divided into $B$ resource blocks (RBs), indexed by $\mathcal{B}=\{1,...,B\}$. We denote by $y_{kb}$ a binary decision variable indicating if RB $b$ is assigned to UE $k$ for the communication with the associated APs, where multiple RBs can be assigned to UE $k$ for task transmission. The channel coefficients on RB $b$ between UE $k$ and AP $j$, UE $k$ and IRS $j$, as well as IRS $j$ and AP $j$, are denoted by $h_{kjb}$, $\boldsymbol{h}_{kjb}^\text{I} = [{h}^\text{I}_1,h^\text{I}_2,...,h^\text{I}_N]$, $\boldsymbol{g}_{jb} = [g_{j1}, g_{j2},..., g_{jN}]^T$, respectively, which are assumed to be perfectly estimated.\footnote{In order to provide some insights and to characterize the performance gain from integrating MEC and IRS, we assume that the channel state information (CSI) for all the wireless links is perfectly known at the AP. Although it is generally hard to obtain perfect CSI, enormous techniques have recently been elaborated to efficiently obtain the CSI for IRS-assisted wireless systems, which can be adopted in our proposed system model to provide accurate CSI \cite{Wei_2021_Channel, taha2019enabling}.} We set the amplitude reflection coefficient to 1 for all IRS reflection elements, and denote the phase shift coefficient vector for IRS $j$ by $\boldsymbol{\Theta}_j = [\theta_{j1},\theta_{j2},...,\theta_{jN}]^T$, where $\theta_{jn} \in [0,2\pi)$ for all $j\in\mathcal{M}, n\in\mathcal{N}$. Then, the reflection-coefficient vector of IRS $j$ is denoted by $\boldsymbol{v}_j = [e^{i\theta_{j1}},e^{i\theta_{j2}},...,e^{i\theta_{jN}}]^T$, where $i$ represents the imaginary unit, and we define $\boldsymbol{\Theta}_j = \text{diag}(\boldsymbol{v}_j)$.

The transmission rate from UE $k$ to AP $j$ on RB $b$, $R_{kjb}(\boldsymbol{\theta}_j)$, is defined as~\blue{\cite{9388935}}:
\begin{equation}
\label{eq:scRate} R_{kjb}(p_k,\boldsymbol{\theta}_j) = W \log_2\left(1+\frac{p_k |h_{kjb} + \boldsymbol{h}_{kjb}^\text{I} \boldsymbol{\Theta}_j \boldsymbol{g}_{jb}|^2}{W N_0}\right)\\
\end{equation}
respectively, where $p_k$ is a decision variable denoting UE $k$'s transmission power on the assigned RBs out of the maximum power $\bar{P}_k$, $W$ is the RB bandwidth, and $N_0$ is the white noise power level. Note that, a large block fading is assumed such as the channel fading is considered constant throughout the duration of the task transmission.
    
\subsection{Latency and Reliability Models}
\blue{By denoting $\boldsymbol{y}_k=\{y_{kb},\forall b\in\mathcal{B}\}$,} the latency incurred from transmitting UE $k$'s task to AP $j$, is given by:
\begin{equation}
\label{eq:latU} L_{kj}^\text{u}(\boldsymbol{y}_k,p_k,\boldsymbol{\theta}_j) = \frac{D_k}{\sum_{b\in\mathcal{B}}y_{kb} R_{kjb}(p_k,\boldsymbol{\theta}_j)}.
\end{equation}

When UE $k$ offloads its task to cloudlet $j$, a portion ($f_{kj}\geq0$) of the computational resources of $j$ ($F_j$) is allocated for the task of UE $k$. Thus, the latency incurred from computing task $k$ on  $j$ depends on the portion $f_{ij}F_j$ of the computational capacity which is allocated and is given by:
\begin{equation}
    \label{eq:latC} L_{kj}^{\text{c}}(f_{kj}) = \frac{D_k C_k}{f_{kj} F_j}
\end{equation}

\blue{We consider heterogeneous edge servers that would belong to third-party operators with non-guaranteed reliability represented by $\phi_j$, due to internal factors that could cause cloudlet nodes to fail, and hence affecting the tasks' computation such~\cite{kherraf2019latency,el2021uav}. In particular, the edge cloudlets' failures would mainly be caused by either software or hardware failures such as hard disk, memory, and RAID controller failures~\cite{vishwanath2010characterizing}. The reliability indicator for each edge server $j$ can be obtained after estimating the occurrence probability of failure scenarios through statistical means based on the cloudlets' historical failure pattern and maintenance records, such as data indicating the mean time between failures and the mean time to repair for each node~\cite{zhang2014venice}. This is identical to how the failure probabilities are usually computed for server nodes in cloud data centers~\cite{ayoubi2016reliable}.}

To guarantee the reliability of each task, redundant computing resources are allocated to each task on failure-independent edge resources, and thus a task may be offloaded to multiple cloudlets concurrently hosted at different base stations and reached through different IRSs. \blue{The reliability constraint for task $k$ is represented by}:
\begin{equation}
\label{cons:reliab} \blue{1-\prod_{j\in\mathcal{M}}\left(1-x_{kj}\phi_j\right) \geq \bar{R}_k}
\end{equation}
    
Intuitively, more redundant resources will provide better computing reliability at the edge to the offloaded tasks; however, this consumes more network resources, both communication and computing resources.

\section{Problem Formulation}
Our objective is to minimize the total UEs' consumed energy subject to latency and reliability constraints of the various tasks, and the network resources capacity by optimizing the task offloading matrix $\boldsymbol{x}=\{x_{kj},\forall k\in\mathcal{K},j\in\mathcal{M}\}$, the RBs association matrix $\boldsymbol{y}=\{y_{kb},\forall k\in\mathcal{K},b\in\mathcal{B}\}$, the UEs' transmit power vector $\boldsymbol{p}=\{p_k,\forall k\in\mathcal{K}\}$, the UEs' computational resources' allocation matrix $\boldsymbol{f}=\{f_{kj},\forall k\in\mathcal{K},j\in\mathcal{M}\}$, and the phase shift coefficient matrix $\boldsymbol{\theta}=\{\theta_{jn},\forall j\in\mathcal{M},n\in\mathcal{N}\}$. The latency and reliability-aware IRS-aided edge computation offloading framework is formulated as an optimization problem, which can be written as follows:
\begin{subequations}
    \label{pr:P1}
    \begin{align}
    \mathcal{P}_1: \underset{\substack{\boldsymbol{x}, \boldsymbol{y}, \boldsymbol{p}, \\ \boldsymbol{f}, \boldsymbol{\theta}, \boldsymbol{\tau}}}\min \quad
    &\label{pr:obj} \sum_{k\in\mathcal{K}}p_k\left(\frac{D_k}{\tau_k}\right)\\
    \text{s.t.}\quad
    &\label{pr:bound} \sum_{b\in\mathcal{B}}y_{kb} R_{kjb}(p_k,\boldsymbol{\theta}_j) \geq x_{kj} \tau_k, \ \forall k,j\\
    &\label{pr:latency} x_{kj}\left(L_{kj}^\text{u}(\boldsymbol{y}_k,p_k,\boldsymbol{\theta}_j) + 					L_{kj}^{\text{c}}(f_{kj})\right) \leq \bar{L}_k\\
    &\label{pr:reliabConv} \blue{\ln(1-\bar{R}_k) \geq \sum_{j \in \mathcal{M}}x_{kj}\ln(1-\phi_j)}\\
    &\label{pr:bandwCap}\sum_{k\in\mathcal{K}} y_{kb} \leq 1, \ b\in\mathcal{B}\\
    &\label{pr:capacity} \sum_{k\in\mathcal{K}} f_{kj} \leq 1, \ \forall j\in\mathcal{M}\\
    &\label{pr:integ} x_{kj}, y_{kb} \in \{0,1\} \notag\\
    &p_k \in [0,\bar{P}_k], f_{kj} \geq 0, \theta_{jn} \in [0,2\pi)
    \end{align}
\end{subequations}
where \eqref{pr:obj} minimizes the total UEs' offloading energy, which is the product of the transmit power and the highest incurred upload latency. Meanwhile, constraint \eqref{pr:bound} sets $\tau_k$ to be the lowest transmission rate for UE $k$ which is being maximized \blue{as part of the objective}. Constraint \eqref{pr:latency} makes sure the total offloading latency of UE $k$'s task on AP $j$ composed of both the upload and computation latencies, respects the UE's latency deadline. \blue{Constraint \eqref{pr:reliabConv} is the linear form of \eqref{cons:reliab} where the transformation steps are detailed in Appendix~\ref{app:convReliab}}. Constraint \eqref{pr:bandwCap} ensures orthogonal radio resource allocation in the APs radio access network. Constraint \eqref{pr:capacity} respects the computational capacity of all cloudlets. Finally, constraint \eqref{pr:integ} is for the integrality and variable bounding conditions. All mathematical symbols used thus far are summarized in Table $\ref{tab:symbols}$.

One can see that problem \eqref{pr:P1} is a non-convex optimization problem due to the coupling with respect to the phase shift $\boldsymbol{\theta}$ and RBs' allocation $\boldsymbol{y}$ as well as the binary task offloading indicator $\boldsymbol{x}$ in \eqref{pr:latency}, the binary association indicator, and the binary resource block indicator. In general, solving problem \eqref{pr:P1} optimally is very difficult, and there are no standard methods for providing such a solution. Thus, in the next sections, we propose a customized suboptimal solution for solving the problem.

\begin{table}
\centering
\renewcommand{\arraystretch}{1.3}
\resizebox{1\linewidth}{!}{
    \begin{tabular}{l|c}
        \hline
        \bfseries Notation & \bfseries Description\\
        \hline
        $\mathcal{M}$ & Set of APs\\
        \hline
        $\mathcal{K}$ & Set of UEs\\
        \hline
        $\mathcal{N}_j$ & Set of IRS $j$'s elements\\
        \hline
        $\mathcal{B}$ & Set of resource blocks\\
        \hline
        $D_k$ & Task input size (Kbits)\\
        \hline
        $C_k$ & Task computational demand (CPU cycles per bit)\\
        \hline
        $L_k$ & Task latency deadline (milliseconds)\\
        \hline
        $R_k$ & Task reliability requirement\\
        \hline
        $\bar{P}_k$ & UEs' transmit power threshold (dBm)\\
        \hline
        $F_j$ & Cloudlet computational capacity (Ghz)\\
        \hline
        $\phi_j$ & Cloudlet reliability\\
        \hline
        $h_{kjb}$ & Channel coefficient between UE $k$ and AP $j$\\
        \hline
        $\boldsymbol{h}_{kjb}^\text{I}$ & Channel coefficient between UE $k$ and IRS $j$\\
        \hline
        $\boldsymbol{g}_{jb}$ & Channel coefficient between IRS $j$ and AP $j$\\
        \hline
        $\boldsymbol{v}_j$ & Reflection-coefficient vector of IRS $j$\\
        \hline
        $g_0$ & Path loss (PL) at $1$ meter\\
        \hline
        $W$ & RB Radio spectrum bandwidth (Mhz)\\
        \hline
        $N_0$ & White noise power level (dBm/Hz)\\
        \hline
        $x_{kj}\in\{0,1\}$ & if UE $k$ is offloading its task to cloudlet $j$\\
        \hline
        $y_{kb}\in\{0,1\}$ & if RB $b$ is assigned to UE $k$\\
        \hline
        $p_k\geq0$ & UE $k$'s transmission power on the assigned RBs\\
        \hline
        $f_{kj}\geq0$ & Portion of cloudled $j$'s resources allocated to task $k$\\
        \hline
        $\boldsymbol{\Theta}_j$ & Phase shift coefficient vector of IRS $j$\\
        \hline
\end{tabular}}
\caption{Notations Used}
\label{tab:symbols}
\end{table}

\section{Solution Approach}\label{sec:solution}
In this section, we propose a suboptimal solution in order to solve problem \eqref{pr:P1}. Our approach is to decouple the offloading decision from the other variables. By first optimizing the offloading decision, the remaining variables can then be optimized aside, and hence, allowing to great reduction in the complexity of the solution.

\subsection{Optimal Offloading}
For the purpose of optimizing the task offloading decisions, \blue{we consider average allocations of the resources where the UEs offload their tasks to the chosen APs through the average obtained rate on all the RBs, i.e. $y_{kb}=1 \ \forall (k,b)\in(\mathcal{K},\mathcal{B})$), and using the direct APs' signal where $\boldsymbol{\theta}$ is omitted. In addition, the offloading is done using an average constant power $\hat{p}_k$, where the computation resources on each AP are equally allocated among the hosted tasks. Thus, equations \eqref{eq:scRate}, \eqref{eq:latU}, and \eqref{eq:latC} can be replaced by:}
\begin{subequations}
    \begin{align}
    \label{eq2:scRate} \blue{\hat{R}_{kjb}} & \blue{=W \log_2\left(1+\frac{\hat{p}_k |h_{kjb}|^2}{W N_0}\right)}\\
    \label{eq2:latU} \blue{\hat{L}_{kj}^\text{u}(\boldsymbol{x}_j)} & \blue{=\frac{D}{\sum_{b\in\mathcal{B}}\hat{R}_{kjb}/\sum_{k'\in\mathcal{K}}x_{k'j}}}\\
    \label{eq2:latC} \blue{\hat{L}_{kj}^{\text{c}}(\boldsymbol{x}_j)} & \blue{=\frac{D_k C_k}{F_j/\sum_{k'\in\mathcal{K}}x_{k'j}}}
    \end{align}
\end{subequations}
Then, the task offloading matrix $\boldsymbol{x}$ is optimized through problem $\mathcal{P}_{a}$ as:
\begin{subequations}
    \label{pr:Pa1}
    \begin{align}
        \mathcal{P}_{a}: \blue{\underset{\substack{\boldsymbol{x}, \boldsymbol{\tau}}}\min \quad}
        &\label{pra:obj} \blue{\sum_{k\in\mathcal{K}}\hat{p}_k\left(\frac{D_k}{\tau_k}\right)}\\
        \text{s.t.}\quad
        &\label{pra:bound} \blue{\frac{\sum_{b\in\mathcal{B}}\hat{R}_{kjb}}{\sum_{k'\in\mathcal{K}}x_{k'j}} \geq x_{kj} \tau_k, \ \forall k,j}\\
        &\label{pra:latency} \blue{x_{kj} \left(\hat{L}_{kj}^\text{u}(\boldsymbol{x}_j) + \hat{L}_{kj}^{\text{c}}(\boldsymbol{x}_j)\right) \leq \bar{L}_k, \ \forall \ k,j}\\
        &\label{pra:reliabConv} \ln(1-\bar{R}_k) \geq \sum_{j \in \mathcal{M}}x_{kj}\ln(1-\phi_j)\\
        &\label{pra:integ} x_{kj} \in \{0,1\}
    \end{align}
\end{subequations}
It can be seen that problem $\mathcal{P}_{a}$ is a mixed-integer convex optimization problem. One solution, to make the problem more tractable and scalable, is to relax the binary condition of $\boldsymbol{x}$ where post-processing steps will be performed for fixing the variables into their binary form. However, this would provide a low-quality solution compared to the binary solution. Thus, we propose a customized matching game algorithm for solving this optimization problem, i.e., $\mathcal{P}_a$. The matching game algorithm is designed such as to solve problem $\mathcal{P}_{a}$ while tackling the exact same objective and while satisfying the same latency and reliability constraints. Hence, it provides an alternative algorithmic approach based on the Game theory to solve problem $\mathcal{P}_{a}$ with a much lower complexity \blue{compared to solving problem $\mathcal{P}_{a}$ directly}.

First, we convert the task offloading optimization problem $\mathcal{P}_a$ to a many-to-one matching game by forming \blue{the set $\mathcal{S} = \bigcup_{i=0}^{l} \binom{M}{i}$ which includes all APs subsets where each subset $s\in\mathcal{S}$ contains up to $l$ elements}, and by defining the matching function $\Omega:\{\mathcal{K},\mathcal{S},\succ_\mathcal{K},\succ_\mathcal{S}\}$. \blue{The constant $l$ would be set to a sufficiently high number representing the maximum number of APs a given UE $k$ can offload to. Since each subset in $\mathcal{S}$ is calculated based on a fixed maximum subset size $l$ and is represented as a sum of binomial coefficients, each of which is a polynomial in $M$, the matching game algorithm hence has a polynomial time complexity.}

\blue{The cost function for UE $k\in\mathcal{K}$ for offloading their task to subset $s\in\mathcal{S}$ which needs to be minimized, is set to be the energy incurred from offloading to the APs $j\in{s}$, and is defined as:}
\begin{equation}
    \label{eq:utilityUE} \blue{C_k(s) = \hat{p}_k \times \max_{j\in{s}} \left(\hat{L}_{kj}^\text{u}(\boldsymbol{x}_j)\right)}
\end{equation}
\blue{which is aligned with the objective \eqref{pra:obj} of problem $\mathcal{P}_{a}$. On the other hand, the cost function for subset $s\in\mathcal{S}$ for hosting the task of UE $k\in\mathcal{K}$ which needs to be minimized, is set to be the maximum total latency incurred from hosting UE $k$'s task on all the APs $j\in{s}$, and is defined as:}
\begin{equation}
    \label{eq:costSubset} \blue{C'_s(k) = \max_{j\in{s}} \left(\hat{L}_{kj}^\text{u}(\boldsymbol{x}_j) + \hat{L}_{kj}^{\text{c}}(\boldsymbol{x}_j)\right)}
\end{equation}
\blue{which is designed for each subset $j\in{s}$ to prefer UEs that minimize the latency impact on all the hosted tasks.}

The pseudocode for the proposed matching game algorithm for solving the offloading decision optimization sub-problem is outlined in Algorithm \ref{alg:matching}. \blue{The algorithm begins by defining the sets of UEs $\mathcal{K}$ and AP subsets $\mathcal{S}$, the achieved subset reliability $R_s \ \forall s\in\mathcal{S}$, and the cost functions $C_k(s)$ and $C'_s(k)$ for the UEs and subsets. Based on $C'_s(k)$, the preference list $\succ_\mathcal{S}$ for each subset $s\in\mathcal{S}$ is established, where the UEs are ordered based on their impact on the latency of the hosted tasks. In each iteration, each unmatched UE $k\in\mathcal{K}$ updates their preference list $\succ_\mathcal{K}$ based on the current UE-subset pairings, excludes the subsets that previously rejected them or cannot meet their latency or reliability requirements, and proposes to their most preferred remaining subset. Then, each subset $s\in\mathcal{S}$ evaluates the UEs' proposals in order of the established preferences $\succ_\mathcal{S}(s)$, and accepts the first UE that can be hosted while respecting the latency requirements of the hosted tasks. The algorithm iterates until no further changes occur in the UE-subset pairings, indicating a stable matching. The final assignment of UEs across the subsets $\Omega$ is then mapped to the task offloading matrix $\boldsymbol{x}$ of problem $\mathcal{P}_{a}$.}

\begin{algorithm}[!t]
    \caption{UEs' Assignment Matching Algorithm}
    \label{alg:matching}
    \begin{algorithmic}[1]
        \State \blue{Define sets $\mathcal{K}, \mathcal{S}$ with $1 < |s| \leq l \ \forall s\in\mathcal{S}$}
        \State \blue{Define sets reliability $R_s \ \forall s\in\mathcal{S}$} 
        \State \blue{Define cost functions $C_k \ \forall k\in\mathcal{K}, C'_s \ \forall s\in\mathcal{S}$}
        \State \blue{Define preference lists $\succ_\mathcal{K} \forall k\in\mathcal{K}, \succ_\mathcal{S} \forall s\in\mathcal{S}$}
        \State \blue{Define matching function $\Omega=\{\Omega(k)\}_{k\in\mathcal{K}}=\varnothing$}
        \State \blue{Define $stable \gets false$} 
        \While{\blue{not stable}}
            \State \blue{$stable \gets true$}
            \For{\blue{each UE $k \in \mathcal{K}$}}
                \If{\blue{$\Omega(k)$ is empty}}
                    \State \blue{Update preference list $\succ_\mathcal{K}(k)$}
                    \State \blue{Perform necessary exclusions from $\succ_\mathcal{K}(k)$}
                    \State \blue{Propose to most preferred subset $s\in\succ_\mathcal{K}(k)$}
                \EndIf
            \EndFor
            \For{\blue{each subset $s\in\mathcal{S}$}}
                \If{\blue{$s$ has proposed UEs}}
                    \State \blue{Sort proposed UEs based on $\succ_\mathcal{S}(s)$}
                    \For{\blue{each proposed UE $k$}}
                        \If{\blue{can accept UE $k$}}
                            \State \blue{$\Omega(k) \gets s$}
                            \State \blue{$stable \gets false$}
                            \State \blue{Break}
                        \EndIf
                    \EndFor
                \EndIf
            \EndFor
        \EndWhile
        \State \blue{Map $\Omega$ to the solution $\boldsymbol{x}^\star$ and return $\boldsymbol{x}^\star$.}
    \end{algorithmic}
\end{algorithm}

\blue{\emph{Convergence Analysis}: The matching game algorithm exhibits convergence through its iterative approach, where in each iteration, UEs dynamically update their preferences and propose to their most preferred subsets based on the current matching, and subsets evaluate these proposals considering their preferences and constraints. After enough iterations, the algorithm reaches convergence when no further changes in matches are observed, signifying a stable state where every UE is paired with a suitable subset, and no UE can improve its assignment by switching to another subset. This stability criterion ensures that the algorithm converges to a stable matching solution in a finite number of steps~\cite{kazmi2017mode}.}

\subsection{IRS Phase Shift Optimization and resources Allocation}
After optimizing the task offloading matrix $\boldsymbol{x}$ through Algorithm \ref{alg:matching}, the RBs association matrix $\boldsymbol{y}$, the UEs' transmit power vector $\boldsymbol{p}$, the UEs' computational resources' allocation matrix $\boldsymbol{f}$, and the IRS phase shift matrix $\boldsymbol{\theta}$, can be optimized through problem $\mathcal{P}_b$ as:
\begin{subequations}
    \label{pr:Pb}
    \begin{align}
        \mathcal{P}_b: \underset{\substack{\boldsymbol{y}, \boldsymbol{p}, \boldsymbol{f}, \\ \boldsymbol{\theta}, \boldsymbol{\tau}}}\min \quad
        &\label{prb:obj} \sum_{k\in\mathcal{K}}p_k\left(\frac{D_k}{\tau_k}\right)\\
        \text{s.t.}\quad
        &\label{prb:bound} \sum_{b\in\mathcal{B}}y_{kb} R_{kjb}(p_k,\boldsymbol{\theta}_j) \geq \tau_k, \ \forall k,j\\
        &\label{prb:latency} L_{kj}^\text{u}(\boldsymbol{y}_k,p_k,\boldsymbol{\theta}_j) + L_{kj}^{\text{c}}(f_{kj}) \leq \bar{L}_k, \ \forall k,j\in\hat{\mathcal{M}}_k\\
        &\label{prb:bandwCap}\sum_{k\in\mathcal{K}} y_{kb} \leq 1, \ b\in\mathcal{B}\\
        &\label{prb:capacity} \sum_{k\in\mathcal{K}} f_{kj} \leq 1, \ \forall j\in\mathcal{M}\\
        &\label{prb:integ} y_{kb} \in \{0,1\} \notag\\
        &p_k \in [0,\bar{P}_k], f_{kj} \geq 0, \theta_{jn} \in [0,2\pi)
    \end{align}
\end{subequations}
where $\hat{\mathcal{M}}_k$ represents the subset of APs that UE $k$ is assigned to, e.g. where $\hat{x}_{kj}=1$. Since problem $\mathcal{P}_b$ is still non-convex, similar to $\mathcal{P}_1$, we approach a suboptimal solution for $\mathcal{P}_b$ by decoupling the optimization of the IRS phase shift from that of the RBs' allocation and the UEs' transmit power. Then, the two sub-problems are iteratively solved until convergence. This decoupling technique will allow for a further reduction of the solution complexity and hence making it more scalable.

\subsubsection{Optimal IRS configuration}
For the given values of the task offloading matrix $\boldsymbol{\hat{x}}$, the RBs' allocation $\boldsymbol{\hat{y}}$, the UEs' transmit power vector $\boldsymbol{\hat{p}}$, and the UEs' computational resources' allocation matrix $\boldsymbol{\hat{f}}$, the IRS phase shift $\boldsymbol{\theta}$ can be optimized as:
\begin{subequations}
    \label{pr:Pc1}
    \begin{align}
    \mathcal{P}_{c1}: \blue{\underset{\substack{\boldsymbol{\theta}, \boldsymbol{\tau}, \boldsymbol{u}}}\min \quad}
    &\label{prc1:obj} \blue{\sum_{k\in\mathcal{K}}\hat{p}_k\left(\frac{D_k}{\tau_k}\right)}\\
    \text{s.t.}\quad
    &\label{prc1:bound} \sum_{b\in\hat{\mathcal{B}}_k}u_{kjb} \geq \tau_k, \ \forall (k,j)\in(\mathcal{K},\hat{\mathcal{M}}_k)\\
    &\label{prc1:rate} R_{kjb}(\hat{p}_k,\boldsymbol{\theta}_j) \geq u_{kjb}, \ \forall (k,j,b)\in(\mathcal{K},\hat{\mathcal{M}}_k,\hat{\mathcal{B}}_k)\\
    &\label{prc1:integ} \theta_{jn} \in [0,2\pi)
    \end{align}
\end{subequations}
where $\hat{\mathcal{B}}_k$ represents the subset of RBs that UE $k$ is assigned to (where $\hat{y}_{kb}=1$), and $u_{kjb}\geq0$ is a newly introduced slack variable to facilitate the transformation of the problem. \blue{The objective \eqref{prc1:obj} guarantees that the IRS elements' phase shifts will be optimized such as to contribute toward minimizing the total UEs' energy consumption when used as the input $\hat{\boldsymbol{\theta}}$ in the resources' allocation optimization problem (problem $\mathcal{P}_{d1}$).}

One can see that the optimization problem $\mathcal{P}_{c1}$ is hard to solve using traditional optimization techniques due to the non-convex constraints in \eqref{prc1:rate}. In order to tackle this challenge, we apply the following steps to efficiently solve the problem at hand. 1) we opt to conic approximation technique to formulate the optimization problem $\mathcal{P}_{c1}$ as a second-order cone problem (SOCP), i.e. $\mathcal{P}_{c2}$. 2) we reformulate problem $\mathcal{P}_{c2}$ into a rank-one constrained optimization problem via change-of-variables and matrix lifting that give us problems $\mathcal{P}_{c3}$ and $\mathcal{P}_{c4}$, respectively. 3) we utilize a difference-of-convex representation for the rank-one constraint to get problem $\mathcal{P}_{c5}$. 4) Moreover, an efficient successive convex approximation is applied to problem $\mathcal{P}_{c5}$, which results in problem $\mathcal{P}_{c6}$.  That being said, a detailed solution approach can be explained in the following paragraphs.

Specifically, we observe that the resulting formulation yields a generalized convex problem owing to the generalized exponential cone constraint \eqref{prc1:rate}, which continues to make the problem hard to solve, particularly when using a generalized convex solver such as FMINCON. On the other hand, a SOCP can be more efficiently solved using MOSEK or any other commercial solver, while attaining a high accuracy of $99.99\%$~\cite{ben2001polyhedral}. Hence, we opt to employ the conic approximation with controlled accuracy in \cite{ben2001polyhedral}, which redefines constraint \eqref{prc1:rate} as a set of the following second-order cone inequalities:
\begin{align}
\label{prFc:main} \kappa_{q+4}^{kjb} & \leq 1 + \left(\hat{p}_k |h_{kjb} + \boldsymbol{h}_{kjb}^\text{I} \boldsymbol{\Theta}_j \boldsymbol{g}_{jb}|^2/W N_0\right)\\
a + \kappa_1^{kjb} & \geq \vectnorm{a - \kappa_1^{kjb} & 2 + u_{kjb}/W 2^{m-1}} \notag\\
a + \kappa_2^{kjb} & \geq \vectnorm{a - \kappa_2^{kjb} & 5/3 + u_{kjb}/W 2^m} \notag\\
\label{prFc:kappa} a + \kappa_3^{kjb} & \geq \vectnorm{a - \kappa_3^{kjb} & 2\kappa_1^{kjb}}\\
\kappa_4^{kjb} & \geq \kappa_2^{kjb} + \kappa_3^{kjb}/24 + 19/72a \notag\\
a + \kappa_l^{kjb} & \geq \vectnorm{a - \kappa_l^{kjb} & 2\kappa_{l-1}^{kjb}} \forall \ l \in \{5, ..., q+3\} \notag\\
a + \kappa_{q+4}^{kjb} & \geq \vectnorm{a - \kappa_{q+4}^{kjb} & 2\kappa_{q+3}^{kjb}} \notag
\end{align}
where $\boldsymbol{\kappa}_m=\{\kappa_m^{kjb}\geq0,k\in\mathcal{K},j\in\hat{\mathcal{M}}_k,b\in\hat{\mathcal{B}}_k\}$, $a$ is a constant with $a=1$, and $q$ is the conic approximation parameter which can be chosen as $q=4$ to attain a high accuracy. By replacing the term $h_{kjb} + \boldsymbol{h}_{kjb}^\text{I} \boldsymbol{\Theta}_j \boldsymbol{g}_{jb}$ in \eqref{prFc:main} by $h_{kjb}+\boldsymbol{v}_j^H\boldsymbol{\Phi}_{kjb}$ with $\boldsymbol{\Phi}_{kjb} = \text{diag}(\boldsymbol{h}_{kjb}^\text{I})\boldsymbol{g}_{jb}$, problem $\mathcal{P}_{c1}$ can be transformed to:
\allowdisplaybreaks
\begingroup
    \begin{subequations}
        \label{pr:prc2}
        \begin{align}
        \mathcal{P}_{c2}: \blue{\underset{\boldsymbol{\theta}, \boldsymbol{\tau}, \boldsymbol{u}}\min \quad}
        &\label{prc2:obj} \blue{\sum_{k\in\mathcal{K}}\hat{p}_k\left(\frac{D_k}{\tau_k}\right)}\\
        \text{s.t.}\quad
        &\label{prc2:main} \kappa_{q+4}^{kjb} \leq 1 + \hat{p}_k/W N_0 \big(\boldsymbol{v}_j^H\boldsymbol{\Phi}_{kjb}\boldsymbol{\Phi}_{kjb}^H\boldsymbol{v}_j\notag\\ 
        & \quad + \boldsymbol{v}_j^H\boldsymbol{\Phi}_{kjb}h_{kjb} + h_{kjb}^H\boldsymbol{\Phi}_{kjb}^H\boldsymbol{v}_j + |h_{kjb}|^2\big)\\
        &\label{prc2:v} |v_{jn}|^2=1, \text{\eqref{prc1:bound}, \eqref{prFc:kappa}}
        \end{align}
    \end{subequations}
\endgroup
We introduce a slack variable $\boldsymbol{t}=\{t_j,\forall j\in\mathcal{M}\}$, thus problem $\mathcal{P}_{c2}$ can be converted to a homogeneous QCQP as:
\begin{subequations}
    \label{pr:prc3}
    \begin{align}
    \mathcal{P}_{c3}: \blue{\underset{\boldsymbol{\theta}, \boldsymbol{\tau}, \boldsymbol{u}}\min \quad}
    &\label{prc3:obj} \blue{\sum_{k\in\mathcal{K}}\hat{p}_k\left(\frac{D_k}{\tau_k}\right)}\\
    \text{s.t.}\quad
    &\label{prc3:main} \kappa_{q+4}^{kjb} \leq 1 + \hat{p}_k/W N_0 \left(\boldsymbol{\bar{v}}_j^H\boldsymbol{R}_{kjb}\boldsymbol{\bar{v}}_j + |h_{kjb}|^2\right)\\
    &\label{prc3:v} |v_n|^2=1, \text{\eqref{prc1:bound}, \eqref{prFc:kappa}}
    \end{align}
\end{subequations}
\[
\boldsymbol{\Psi}_{kjb}=
\begin{bmatrix}
\boldsymbol{\Phi}_{kjb}\boldsymbol{\Phi}_{kjb}^H & h_{kjb}\boldsymbol{\Phi}_{kjb}\\
h_{kjb}\boldsymbol{\Phi}_{kjb}^H & 0
\end{bmatrix}\forall j,
\ \boldsymbol{\bar{v}}_j=
\begin{bmatrix}
\boldsymbol{v}_j\\
t_j
\end{bmatrix}
\]
    
We note that $\boldsymbol{\bar{v}}_j^H\boldsymbol{\Psi}_{kjb}\boldsymbol{\bar{v}}_j = \text{tr}(\boldsymbol{\Psi}_{kjb}\boldsymbol{V}_j)$ with $\boldsymbol{V}_j=\boldsymbol{\bar{v}}_j\boldsymbol{\bar{v}}_j^T$. Then, problem $\mathcal{P}_{c3}$ can be transformed to:
\begin{subequations}
    \label{pr:prFc4}
    \begin{align}
        \mathcal{P}_{c4}: \blue{\underset{\boldsymbol{V}, \boldsymbol{\tau}, \boldsymbol{u}}\min \quad}
        &\label{prFc4:obj} \blue{\sum_{k\in\mathcal{K}}\hat{p}_k\left(\frac{D_k}{\tau_k}\right)}\\
        \text{s.t.}\quad
        &\label{prFc4:main} \kappa_{q+4}^{kjb} \leq 1 + \hat{p}_k/W N_0 \left(\boldsymbol{\bar{v}}_j^H\boldsymbol{\Psi}_{kjb}\boldsymbol{\bar{v}}_j + |h_{kjb}|^2\right)\\
        &\label{prFc4:v} V_{j,n,n}=1, \ \forall n\in\{1,2,...,N+1\}\\
        &\label{prFc4:SDR} \boldsymbol{V}_j\succcurlyeq0\\
        &\label{prFc4:rank} \text{rank}(\boldsymbol{V}_j)=1, \ \text{\eqref{prc1:bound}, \eqref{prFc:kappa}}
    \end{align}
\end{subequations}

The above problem is a non-convex semi-definite program (SDP) where constraints \eqref{prFc4:SDR} and \eqref{prFc4:rank} indicate that $\boldsymbol{V}_j$ is a semi-definite matrix with a non-convex rank-one constraint. Such a property of the matrix makes it quite hard to solve the above optimization problem. Indeed, semi-definite relaxation (SDR)~\cite{so2007approximating} (which is very popular and widely used to solve similar problems), can be leveraged. In this method, the rank-one constraint is relaxed and this converts the problem into a convex SDP that can be easily solved, and the Gaussian randomization is used to construct a rank-one solution from $\boldsymbol{V}_j$~\cite{so2007approximating}. Nonetheless, such a technique does not guarantee to find a feasible (or a rank-one) solution to \eqref{pr:prFc4}, particularly for larger size problems. As a result, we present an approach for solving \eqref{pr:prFc4} using successive convex approximation (SCA) with convergence guarantees, where the rank-one constraint is presented in a Difference of Convex (DC) form \cite{10225434}.
    
Let $\sigma_i(\boldsymbol{V}_j)$ be the $i$-th largest singular value of $\boldsymbol{V}_j$, then the rank-one constraint for $\boldsymbol{V}_j$ shows that $\sigma_1(\boldsymbol{V}_j)>0$ and $\sigma_i(\boldsymbol{V}_j)=0, \forall i$. Thus, by defining the trace and spectral norms of $\boldsymbol{V}_j$ as $\text{Tr}(\boldsymbol{V}_j)=\sum_{n=1}^{N+1}\sigma_i(\boldsymbol{V}_j)$ and $||\boldsymbol{V}_j||=\sigma_1(\boldsymbol{V}_j)$, respectively, the rank-one constraint can be rewritten as the difference of the convex norms as $\sum_{j\in\mathcal{M}}\left(\text{Tr}(\boldsymbol{V}_j)-||\boldsymbol{V}_j||\right)=0$ with $\text{Tr}(\boldsymbol{V}_j)>0$. Accordingly, $\mathcal{P}_{c4}$ is re-written in the following DC representation:

\begin{subequations}
    \label{pr:prFc5}
    \begin{align}
        \mathcal{P}_{c5}: \underset{\boldsymbol{V}, \boldsymbol{\tau}, \boldsymbol{u}}\min \quad
        &\label{prFc5:obj} \blue{\sum_{k\in\mathcal{K}}\hat{p}_k\left(\frac{D_k}{\tau_k}\right)} + \sum_{j\in\mathcal{M}}\left(\text{Tr}(\boldsymbol{V}_j)-||\boldsymbol{V}_j||\right)\\
        \text{s.t.}\quad
        &\label{prFc5:remain} \text{\eqref{prc1:bound}, \eqref{prFc:kappa}, \eqref{prFc4:main}, \eqref{prFc4:v}}
    \end{align}
\end{subequations}

The above is still non-convex owing to a concave term $-||\boldsymbol{V}_j||$ in \eqref{prFc5:obj}. 
Hence, we linearize the term $f(\boldsymbol{V}_j)=||\boldsymbol{V}_j||$ by replacing it with $\tilde{f}(\boldsymbol{V}_j;\boldsymbol{V}_j^{(n)})=\Big\langle\psi||\boldsymbol{V}_j^{(n)}||,\boldsymbol{V}_j\Big\rangle$, where $\boldsymbol{V}_j^{(n)}$ is the solution obtained at iteration $n$ of the SCA method, and the sub-gradient of spectral norm at point $\boldsymbol{V}_j^{(n)}$ is denoted by $\psi||\boldsymbol{V}_j^{(n)}||$. Please note that one sub-gradient of $||\boldsymbol{V}_j||$ can be effectively evaluated as $\boldsymbol{q}_1 \boldsymbol{q}_1^H$, where $\boldsymbol{q}_1$ is the vector corresponding to the largest singular value $\sigma_1(\boldsymbol{V}_j)$. Thus, $\mathcal{P}_{c5}$ can be rewritten as:
\begin{subequations}
    \label{pr:prFc6}
    \begin{align}
        \mathcal{P}_{c6}^{(n)}: \underset{\boldsymbol{V}, \boldsymbol{\tau}, \boldsymbol{u}}\min \quad
        &\label{prFc6:obj} \Gamma_c = \blue{\sum_{k\in\mathcal{K}}\hat{p}_k\left(\frac{D_k}{\tau_k}\right)}\notag \\
        &\quad+ \sum_{j\in\mathcal{M}}\left(\text{Tr}(\boldsymbol{V}_j)-\tilde{f}(\boldsymbol{V}_j;\boldsymbol{V}_j^{(n)})\right)\\
        \text{s.t.}\quad
        &\label{prFc6:remain} \text{\eqref{prc1:bound}, \eqref{prFc:kappa}, \eqref{prFc4:main}, \eqref{prFc4:v}}
    \end{align}
\end{subequations}
 This problem $\mathcal{P}_{c6}^{(n)}$ is a standard convex SDP solved at SCA iteration $n$ until convergence, assuming an initial $\boldsymbol{V}_j^{(0)}$. In particular, we utilize a stopping criterion that ensures $\sum_{j\in\mathcal{M}}\left(\text{Tr}(\boldsymbol{V}_j)-\tilde{f}(\boldsymbol{V}_j;\boldsymbol{V}_j^{(n)})\right)<\epsilon$, where $\epsilon$ is a sufficiently small positive constant. After obtaining a feasible $\boldsymbol{V}_j, \forall j\in\mathcal{M}$, the phase shift matrix $\boldsymbol{\theta}_j$ can then be easily retrieved from $\boldsymbol{V}_j, \forall j\in\mathcal{M}$. The algorithm pseudo-code for solving the IRS phase shift optimization sub-problem at each iteration of the alternating optimization, is outlined in Algorithm \ref{alg:subproblemC}.
\begin{algorithm}[!t]
    \caption{IRS Optimization}
    \label{alg:subproblemC}
    \begin{algorithmic}[1]
        \State $\textbf{Initialize:}$ \label{algC:init}
        \State $\boldsymbol{\hat{x}}$, $\boldsymbol{\hat{y}}$, $\boldsymbol{\hat{p}}$, $\boldsymbol{\hat{f}}$, and $n = 0$; \label{algC:define_n}
        \State Select an initial point $\boldsymbol{V}^{(n)}$; \label{algC:define_P0}
        \Repeat \label{algC:while_S}
            \State Solve $\mathcal{P}_{c6}^{(n)}$ to reach an optimal solution $\boldsymbol{\omega}_c^{(n)}=\{\boldsymbol{V}^\star, \boldsymbol{\tau}^\star, \boldsymbol{u}^\star\}$ and objective $\Gamma_c^{(n)}$ at the $n$th iteration. \label{algC:solve}
            \State Update $\boldsymbol{V}^{(n)}=\boldsymbol{V}^\star$; \label{algC:assign_Pn}
            \State $n = n + 1$; \label{algC:increment}
        \Until Convergence (of the objective of $\mathcal{P}_{c6}^{(n)}$). \label{algC:while_E}
        \State Recover the phase shift matrix $\boldsymbol{\theta}^\star$ from $\boldsymbol{V}^\star$.
        \State Return the solution $\boldsymbol{\theta}^\star$ and the objective $\Gamma_c^\star=\Gamma_c^{(n)}$.
    \end{algorithmic}
\end{algorithm}
    
\emph{Convergence Analysis}: Algorithm~\ref{alg:subproblemC} is proved to converge by showing the series of resulting objectives is monotonically convergent. Due to the convex approximation for $\tilde{f}(\boldsymbol{V}_j;\boldsymbol{V}_j^{(n)})$, the updating rules in Algorithm \ref{alg:subproblemC}, c.f., Step \ref{algC:assign_Pn}, ensure that the solution set $\boldsymbol{\omega}_c^{(n)}$ is a feasible solution to problem $\mathcal{P}_{c5}$ at step $n+1$. This subsequently leads to the results of $\Gamma_c^{(n+1)} \leq \Gamma_c^{(n)}$, which means that Algorithm \ref{alg:subproblemC} generates a non-increasing sequence of objective function values. Due to constraint \eqref{prFc4:main}, the sequence of $\Gamma_c^{(n)},n=1,2,\dots$ is bounded below and therefore, Algorithm \ref{alg:subproblemC} guarantees that the objective converges.
    
\subsubsection{Resources Allocation Optimization}
For the given values of the phase shift $\hat{\boldsymbol{\theta}}$, the RBs' allocation $\boldsymbol{y}$, the UEs' transmit power $\boldsymbol{p}$ and the UEs' computational resources' allocation $\boldsymbol{f}$, can be optimized as:
\begin{subequations}
    \label{pr:Pd1}
    \begin{align}
        \mathcal{P}_{d1}: \underset{\substack{\boldsymbol{y}, \boldsymbol{p}, \\ \boldsymbol{f}, \boldsymbol{u}}}\min \quad
        &\label{prd1:obj} \sum_{k\in\mathcal{K}}p_k\left(\frac{D_k}{\tau_k}\right)\\
        \text{s.t.}\quad
        &\label{prd1:bound} \sum_{b\in\mathcal{B}}u_{kjb} \geq \tau_k, \ \forall (k,j)\in(\mathcal{K},\hat{\mathcal{M}}_k)\\
        &\label{prd1:rate} y_{kb} R_{kjb}(p_k,\boldsymbol{\hat{\theta}}_j) \geq u_{kjb}, \ \forall (k,j,b)\in(\mathcal{K},\hat{\mathcal{M}}_k,\mathcal{B})\\
        &\label{prd1:latency} L_{kj}^\text{u}(\boldsymbol{y}_k,p_k,\boldsymbol{\hat{\theta}}_j) + L_{kj}^{\text{c}}(f_{kj}) \leq \bar{L}_k, \ \forall j\in\hat{\mathcal{M}}_k\\
        &\text{\eqref{prb:bandwCap}, \eqref{prb:capacity}}\\
        &\label{prd1:integ} y_{kb} \in \{0,1\}, p_k \in [0,\bar{P}_k], f_{kj} \geq 0
    \end{align}
\end{subequations}
Problem $\mathcal{P}_{c}$ is a mixed-integer non-convex program due to the binary condition of variable $\boldsymbol{y}$ in \eqref{prd1:integ}, and to \eqref{prd1:obj} being neither convex nor concave in general. We note that the generalized exponential cone constraint \eqref{prd1:rate} can be rewritten into a set of second-order cone inequalities similar to \eqref{prc1:rate} with $a=y_{kb}$, and the generalized convex constraint \eqref{prd1:latency} can be easily transformed into a second-order cone constraint. In order to convexify problem $\mathcal{P}_{c}$, we introduce slack variable $\delta_k, \forall k\in\mathcal{K}$ which will equivalently replace \eqref{prd1:obj}, and slack variable $\Lambda_k, \forall k\in\mathcal{K}$, and equivalently define the following constraints:
\begin{subequations}
    \begin{align}
    \label{eq:appr1} \delta_k \tau_k \geq \Lambda_k^2\\
    \label{eq:appr2} p_k - \Lambda_k^2 \leq 0
    \end{align}
\end{subequations}
Here, \eqref{eq:appr1} is a quadratic conic convex constraint, while \eqref{eq:appr2} is non-convex due to the concave function $g(\Lambda_k)=-\Lambda_k^2$ which renders the left side of \eqref{eq:appr2} as a DC form. Thus, with $\Lambda^{(m)}$ as the input point, we use the SCA method to substitute $g(\Lambda)$ by its first order Taylor approximate as:
\begin{align}
\tilde{g}(\Lambda_k;\Lambda_k^{(m)}) = - (\Lambda_k^{(m)})^2 - 2\Lambda_k^{(m)}(\Lambda_k-\Lambda_k^{(m)}) \label{approx:2}
\end{align}
The mixed-integer approximation of $\mathcal{P}_{d1}$ continues to be not scalable, making the SCA algorithm not suitable for big instances due to the mixed-integer nature of the problem, which is caused by the binary condition of variable $\boldsymbol{y}$ in \eqref{prd2:integ}. To deal with this, we adopt a similar approach to \cite{nguyen2018novel}, where we relax the binary condition for variable $\boldsymbol{y}$ by introducing the following constraint:
\begin{align}
\label{prd2:xAppr} 0 \leq y_{kb} - \tilde{g}(y_{kb}; y_{kb}^{(m)}) \leq \zeta_{kb}
\end{align}
which is the approximated form of the original non-convex constraint, where $\boldsymbol{\zeta}=\{\zeta_{kb}\geq0,\forall k\in\mathcal{K},b\in\mathcal{B}\}$ is a newly introduced slack variable. Constraint \eqref{prd2:xAppr} will force variable $\boldsymbol{y}$ to take a binary value with a penalty term added to the objective. At this point, problem $\mathcal{P}_{d1}$ can then be replaced by:
\begin{subequations}
    \label{pr:Pd2}
    \begin{align}
    \mathcal{P}_{d2}^{(m)}: \underset{\substack{\boldsymbol{y}, \boldsymbol{p}, \boldsymbol{f}, \\ \boldsymbol{u}, \boldsymbol{\Lambda}}}\min \quad
    &\label{prd2:obj} \Gamma_d = \sum_{k\in\mathcal{K}}D_k \delta_k + A\sum_{k\in\mathcal{K}}\sum_{b\in\mathcal{B}}\zeta_{kb}\\
    \text{s.t.}\quad
    &\label{prd2:bound} \sum_{b\in\mathcal{B}}u_{kjb} \geq \tau_k, \ \forall (k,j)\in(\mathcal{K},\hat{\mathcal{M}}_k)\\
    &\label{prd2:rate} \kappa_{q+4}^{kjb} \leq y(k,b) + \notag\\ & \qquad \left(p_k |h_{kjb} + \boldsymbol{h}_{kjb}^\text{I} \hat{\boldsymbol{\Theta}}_j \boldsymbol{g}_{jb}|^2/W N_0\right)\\
    &\label{prd2:latency} L_{kj}^\text{u}(\boldsymbol{y}_k,p_k,\boldsymbol{\hat{\theta}}_j) + L_{kj}^{\text{c}}(f_{kj}) \leq \bar{L}_k, \ \forall j\in\hat{\mathcal{M}}_k\\
    &\label{prd2:appr2} p_k + \tilde{g}(\Lambda_k;\Lambda_k^{(m)}) \leq 0\\
    &\text{\eqref{prb:bandwCap}, \eqref{prb:capacity}, \eqref{prFc:kappa}, \eqref{eq:appr1}, \eqref{prd2:xAppr}}\\
    &\label{prd2:integ} y_{kb} \in [0,1], p_k \in [0,\bar{P}_k], f_{kj} \geq 0
    \end{align}
\end{subequations}
The problem is now a SOCP to be solved at each iteration $m$ until convergence, with $A>0$ being the penalty parameter for reinforcing the binary condition on variable $\boldsymbol{y}$. Here, $\boldsymbol{\Lambda}^{(m)}$ and $\boldsymbol{y}^{(m)}$ are being updated to the optimal $\boldsymbol{\Lambda}$ and $\boldsymbol{y}$, respectively, at each iteration of the SCA-based algorithm. The algorithm pseudocode, for solving the resources' allocation optimization sub-problem at each iteration of the alternating optimization, is outlined in Algorithm \ref{alg:subproblemD}.
\begin{algorithm}[!t]
    \caption{Resources Allocation Optimization}
    \label{alg:subproblemD}
    \begin{algorithmic}[1]
        \State $\textbf{Initialize:}$ \label{algD:init}
        \State $\boldsymbol{\hat{x}}$, $\boldsymbol{\hat{\theta}}$, and $m = 0$; \label{algD:define_n}
        \State Select an initial point $\boldsymbol{\Lambda}^{(m)}$; \label{algD:define_P0}
        \Repeat \label{algD:while_S}
            \State Solve $\mathcal{P}_{d2}^{(m)}$ to reach an optimal solution $\boldsymbol{\omega}_d^{(m)}=\{\boldsymbol{y}^\star, \boldsymbol{p}^\star, \boldsymbol{f}^\star, \boldsymbol{u}^\star, \boldsymbol{\Lambda}^\star\}$ and objective $\Gamma_d^{(m)}$ at the $m$th iteration. \label{algD:solve}
            \State Update $\boldsymbol{\Lambda}^{(m)}=\boldsymbol{\Lambda}^\star$; \label{algD:assign_Pn}
            \State $m = m + 1$; \label{algD:increment}
        \Until Convergence of the objective of $\mathcal{P}_{d2}^{(m)}$. \label{algD:while_E}
        \State Return the solution $\{\boldsymbol{y}^\star,\boldsymbol{p}^\star,\boldsymbol{f}^\star\}$ and the objective $\Gamma_d^\star=\Gamma_d^{(m)}$.
    \end{algorithmic}
\end{algorithm}

\emph{Convergence Analysis of Alg.~\ref{alg:subproblemD}}: Algorithm~\ref{alg:subproblemD} is shown to converge by showing the series of resulting objectives is monotonically convergent. Owing to a convex approximation in \eqref{approx:2}, the updating rules in Algorithm \ref{alg:subproblemD} (c.f., Step \ref{algD:assign_Pn}) ensure that the solution set $\boldsymbol{\omega}_d^{(m)}$ is a feasible one to $\mathcal{P}_{d1}$ at step $n+1$. This subsequently leads to the results of $\Gamma_d^{(m+1)} \leq \Gamma_d^{(m)}$, which means that Algorithm \ref{alg:subproblemD} generates a non-increasing sequence of objective function values. Due to the latency constraints in \eqref{prd2:latency}, the sequence of $\Gamma_d^{(m)},m=1,2,\dots$ is bounded below and therefore, Algorithm \ref{alg:subproblemD} guarantees that the objective converges. The overall algorithm pseudocode for providing an efficient suboptimal solution for problem $\mathcal{P}_1$, is outlined in Algorithm \ref{alg:problem}.

\begin{algorithm}[!t]
    \caption{Suboptimal Solution for Solving $\mathcal{P}_1$}
    \label{alg:problem}
    \begin{algorithmic}[1]
        \State Obtain $\boldsymbol{x}^\star$ through Algorithm \ref{alg:matching}.
        \State Start with initial values: $\boldsymbol{y}^\star=1$, $\boldsymbol{p}^\star=\boldsymbol{\bar{P}}$, and $\boldsymbol{f}^\star=1$;
        \State Set $i = 1$
        \Repeat
            \State Obtain $\boldsymbol{\theta}^{(i)}$ and the objective $\Gamma_c^{(i)}$ through Algorithm \ref{alg:subproblemC} given the inputs $\boldsymbol{\hat{y}}=\boldsymbol{y}^{(i)}$, $\boldsymbol{\hat{p}}=\boldsymbol{p}^{(i)}$, and $\boldsymbol{\hat{f}}=\boldsymbol{f}^{(i)}$.
            \State Obtain $\{\boldsymbol{y}^{(i)},\boldsymbol{p}^{(i)},\boldsymbol{f}^{(i)}\}$ and the objective $\Gamma_d^{(i)}$ through Algorithm \ref{alg:subproblemD} given the input $\boldsymbol{\hat{\theta}}=\boldsymbol{\theta}^{(i)}$.
            \State Set $i = i + 1$;
        \Until convergence of objective $\Gamma_d^{(i)}$.
        \State Return the optimized solution $\boldsymbol{\omega}^\star=\{\boldsymbol{x}^{(i)},\boldsymbol{y}^{(i)},\boldsymbol{p}^{(i)},\boldsymbol{f}^{(i)},\boldsymbol{\theta}^{(i)}\}$ and objective $\Gamma_d^{(i)}$.
    \end{algorithmic}
\end{algorithm}

\emph{Convergence Analysis Alg.~\ref{alg:problem}}: Algorithm~\ref{alg:problem} is shown to converge by showing that the series of resulting objective is monotonically convergent. Due to the non-increasing sequence of objective function values in Algorithm \ref{alg:subproblemC} and Algorithm \ref{alg:subproblemD}, the objective values $\Gamma_c^{(n)}$ and $\Gamma_d^{(n)}$ at step $i+1$ are guaranteed to be less than or equal to $\Gamma_c^{(n)}$ and $\Gamma_d^{(n)}$ at step $i$, which means that Algorithm \ref{alg:problem} generates a non-increasing sequence of objective function values $\Gamma_d^{(n)}$, which guarantees the objective convergence.

\emph{Complexity Analysis}: The overall complexity of Algorithm \ref{alg:problem} depends mainly on that of solving the SDP problem \eqref{pr:prFc6} in Algorithm \ref{alg:subproblemC}, and the SOCP problem \eqref{pr:Pd2} in Algorithm \ref{alg:subproblemD}. Since the complexity of \eqref{pr:Pd2} is approximately $\mathcal{O}\left(qKMB\right)$ which is a polynomial time complexity with $q$ as the conic approximation parameter~\cite{ben2001polyhedral}, the complexity of solving \eqref{pr:prFc6} is the dominant one. The order of complexity for an SDP problem with $m$ SDP constraints, which includes an $n \times n$ positive semi-define (PSD) matrix, is given by $\mathcal{O}\left(\sqrt{n}\log(1/\epsilon)(m n^3 + m^2 n^2 + m^3)\right)$ where $\epsilon > 0$ is the solution accuracy~\cite[Th.~3.12]{polik2010interior}. In our case, we have $n=N+1$ and $m=KM+qKMB+N+1$; but since $qKMB$ is the dominant term, the approximate computational complexity for solving \eqref{pr:prFc6} (and Algorithm \ref{alg:problem}), can be defined as $\mathcal{O}\left(\log(1/\epsilon)(qKMB N^3 + (qKMB)^2 N^2 + (qKMB)^3)\right)$.

\begin{table}
    \centering
    \renewcommand{\arraystretch}{1.3}
    \resizebox{1\linewidth}{!}{
        \begin{tabular}{l|c}
            \hline
            \bfseries Parameter & \bfseries Value\\
            \hline
            Task input size $D_k$ & $50$ kilobits\\
            \hline
            Task computational demand $C_k$ & $200$ cycles/bit\\
            \hline
            UEs' transmit power threshold $\bar{P}_k$ & $P=30$ dBm~\cite{li2021energy}\\
            \hline
            Cloudlet capacity $F_j$ & $10$ GHz\\
            \hline
            Cloudlet reliability $\phi_j$ & $0.9955$\\
            \hline
            Rician Factor & $2$\\
            \hline
            Path loss (PL) at $1$ meter $g_0$ & $-30$ dB\cite{chu2020intelligent}\\
            \hline
            PL exponents for the UE-AP link (IRS link) & $4$ ($2.2$)\\
            \hline
            RB Radio spectrum bandwidth $W$ & $1$ MHz~\cite{li2021energy}\\
            \hline
            White noise power level $N_0$ & $-174$ dBm/Hz~\cite{li2021energy}\\
            \hline
    \end{tabular}}
    \caption{Instance Parameters}
    \label{tab:params}
\end{table}

\begin{figure*}[!t]
    \centering
    \begin{subfigure}[b]{.495\linewidth}
        \centering
        \includegraphics[width=1\textwidth]{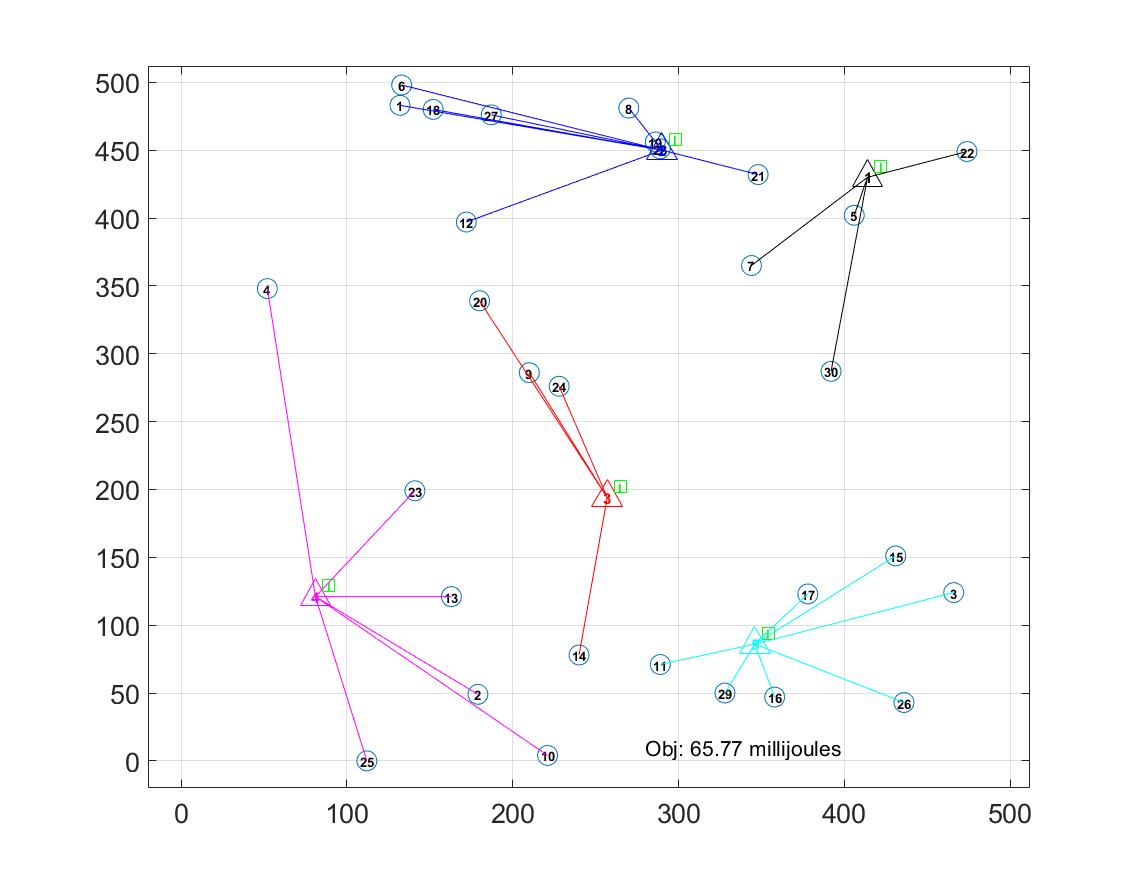}
        \caption{\blue{UEs offloading to the best AP, which alone satisfies their reliability requirement.}}
        \label{fig:plot1Gamea}
    \end{subfigure}
    \begin{subfigure}[b]{.495\linewidth}
        \centering
        \includegraphics[width=1\textwidth]{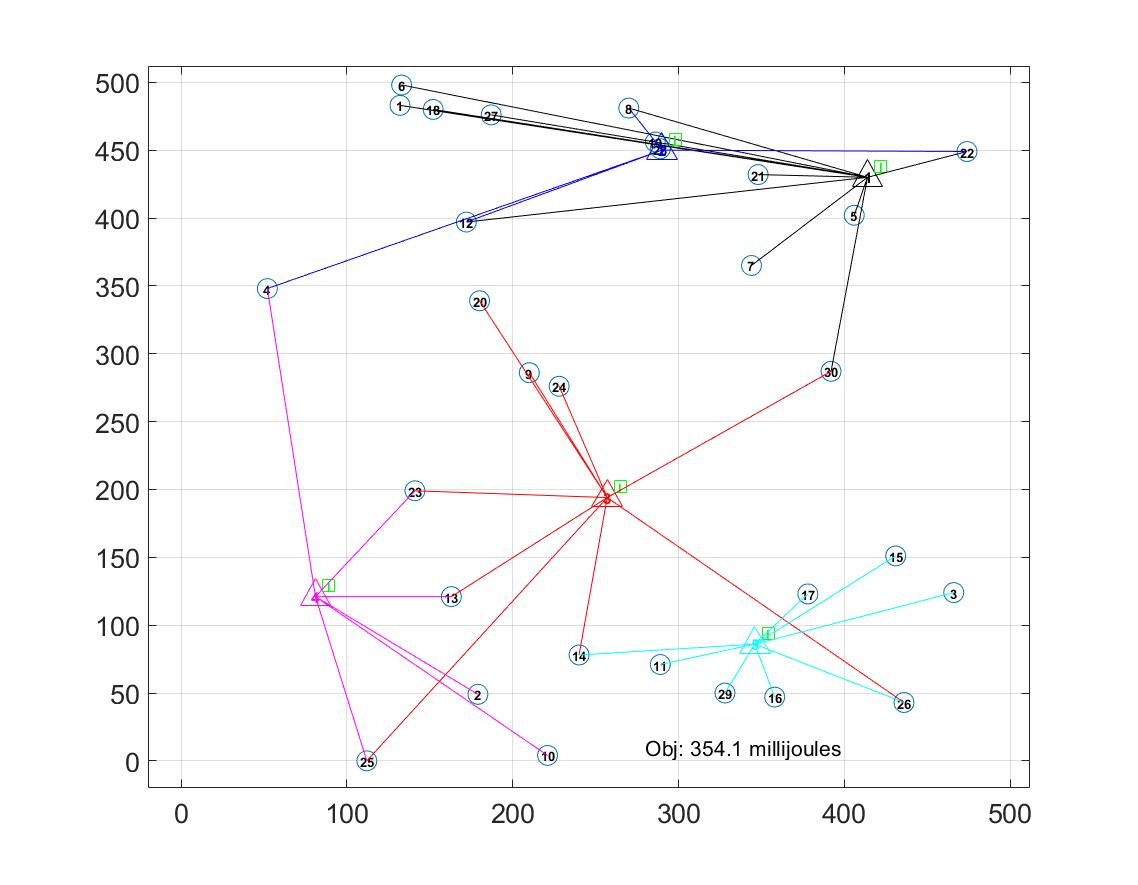}
        \caption{\blue{Some UEs offloading to multiple APs, due to their increased reliability requirement.}}
        \label{fig:plot1Gameb}
    \end{subfigure}
    \caption{Matching game UEs' offloading decisions results. In this figure, triangle, circle, and square represent AP, UE, and IRS, respectively.}
    \label{figs:plot1Game}
\end{figure*}

\section{Numerical Results}\label{sec:results}
The performance of the proposed methods is studied in this section through simulations. The main instance consists of $K=30$ UEs that are placed in the center of a $2$-D area of $400$ m$^2$, within the range of $M=5$ randomly distributed APs, each having a nearby IRS with $N=30$ elements. The system parameters are presented in Table \ref{tab:params}.

\blue{In Fig. \ref{figs:plot1Game}, we show the results obtained from Algorithm \ref{alg:matching}, where the UEs aim at minimizing their offloading energy. In Fig. \ref{fig:plot1Gamea}, all the UEs have reliability requirement $\bar{R}_k = 0.9955, \forall k\in\mathcal{K}$, and hence require to offload their task to only one AP which minimizes their offloading energy (since $\phi_j = 0.9955, \forall j\in\mathcal{M}$). However, in Fig. \ref{fig:plot1Gameb}, the reliability for $10$ of the UEs is set to $\bar{R}_k = 0.9999$, which reflects their need to offload to $2$ APs in order to satisfy their reliability requirement as can be seen in the figure. Here, some UEs such as UE $6$ and $21$ have switched their AP assignment, which is due to the fact that with increased demands on the APs, some UEs can achieve a better transmission rate by switching to other APs which can also guarantee their latency requirement.}

\begin{figure}[!t]
    \centering
    \includegraphics[width=.5\textwidth]{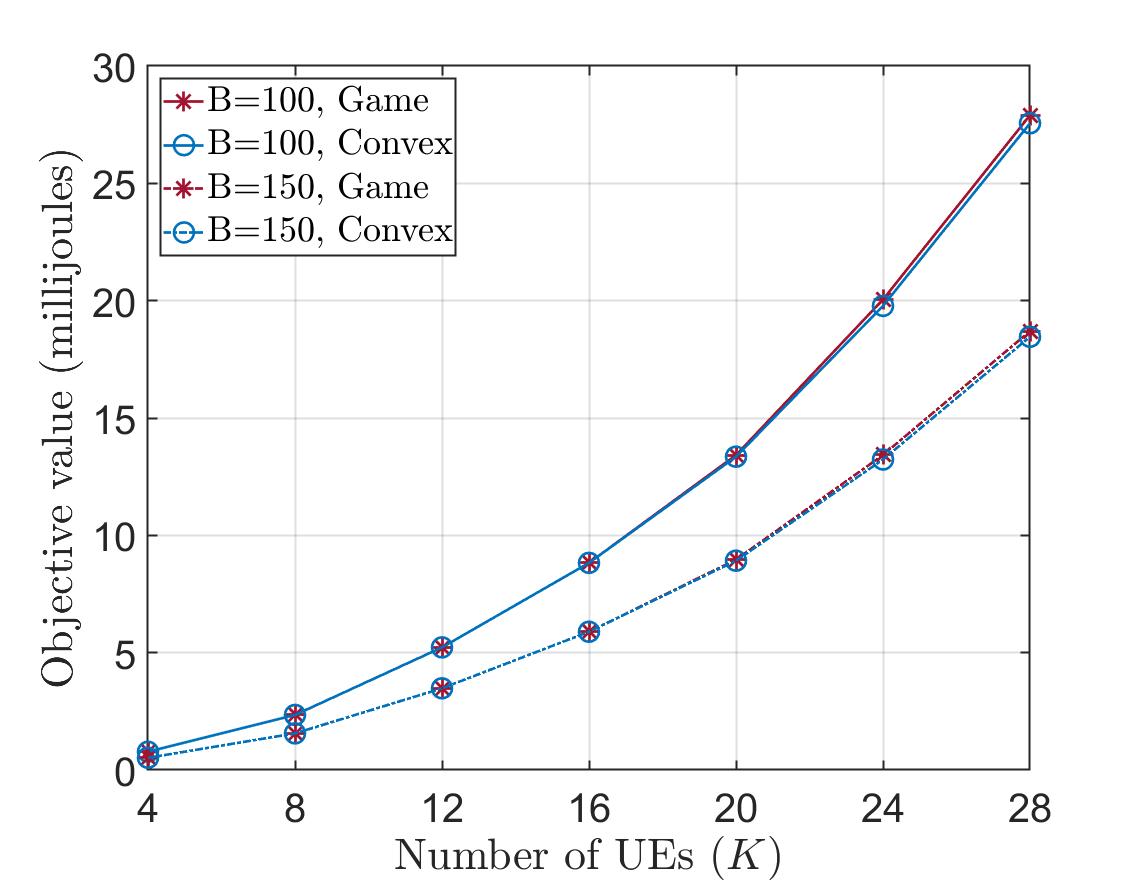}
    \caption{Matching game compared to other approaches.}
    \label{fig:plot5GameComp}
\end{figure}
\blue{In Fig. \ref{fig:plot5GameComp}, we compare the objective value obtained from the matching game algorithm (representing the sum of UEs' offloading energies in millijoules), with that obtained from the solution of the mixed-integer convex problem $\mathcal{P}_{a}$. As it can be seen, the matching game algorithm achieves almost the optimal solution for small values of $K$ with a negligible gap. Then, the gap between the two objective values slowly widens when more UEs are added to the network. It is worth noting here that while achieving an objective that is very close to the optimal value, the matching game algorithm has a much lower complexity and is able to execute large instances very quickly.}

\begin{figure}[!t]
    \centering
    \includegraphics[width=.5\textwidth]{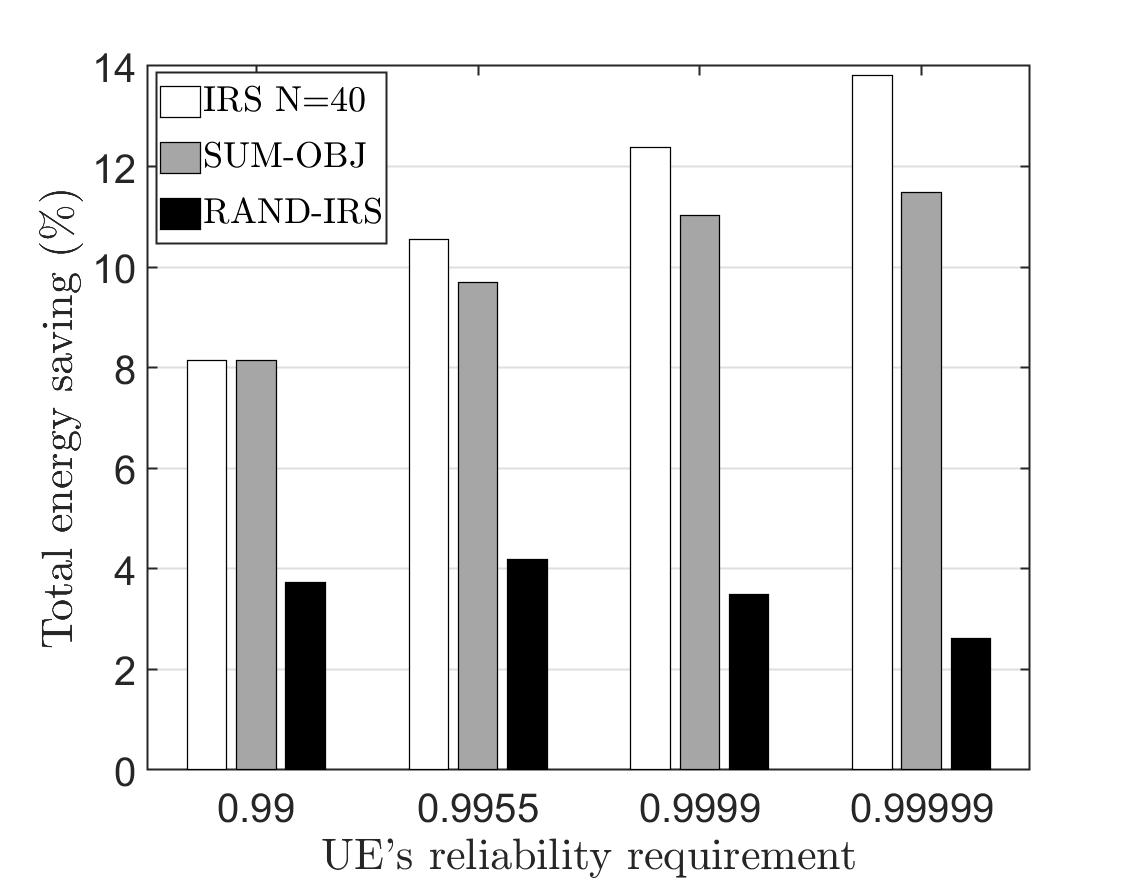}
    \caption{Energy gain incurred from the use of the IRS}
    \label{fig:plot3CompRel}
\end{figure}
Fig. \ref{fig:plot3CompRel} presents the total energy savings of UEs achieved through using IRS as per our solution, and compare it to two approaches: 1) \blue{SUM-OBJ where the objective of problem \eqref{pr:prFc6} is modified such as to maximize the sum of the UEs' transmission rates, i.e. $\underset{\substack{\boldsymbol{\theta}}}\max\sum_{b\in\hat{\mathcal{B}}_k}R_{kjb}(\hat{p}_k,\boldsymbol{\theta}_j)$}, and 2) RAND-IRS that randomly configures the IRS phase shifts. As it can be seen, the proposed solution with $N=40$ outperforms both of SUM-OBJ and RAND-IRS, where the gain over SUM-OBJ is nonexistent in the case of one communication channel, and grows larger with utilizing more APs. It is also observed that the energy saving becomes more significant while increasing the reliability requirement of the UEs. This is because when more APs are utilized, the IRSs can better improve the channels with lower quality, which reduces the UEs' transmission power and therefore their consumed energy.

\begin{figure}[!t]
    \centering
    \includegraphics[width=.5\textwidth]{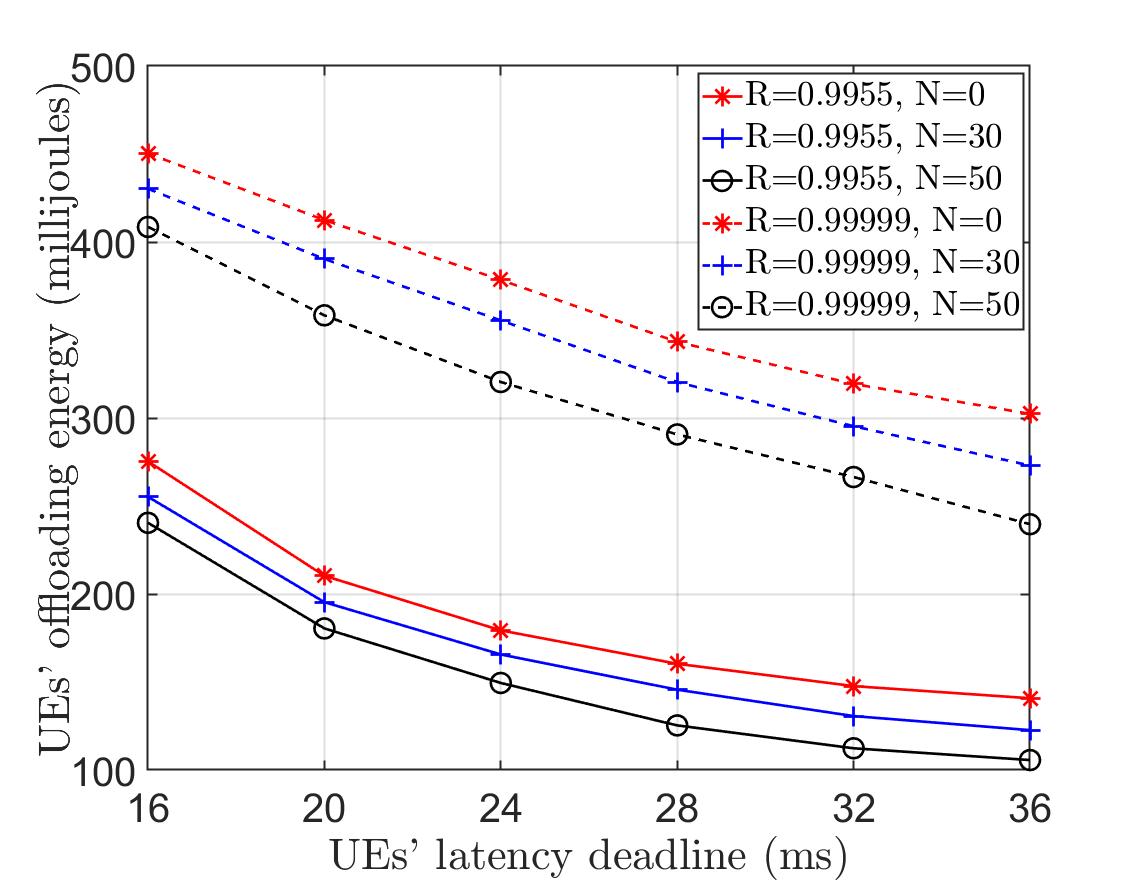}
    \caption{Objective value vs reliability requirement and number of IRS elements}
    \label{fig:plot2LatRelN}
\end{figure}
In Fig. \ref{fig:plot2LatRelN}, we study how the IRS size influences the achieved total energy consumption for different classes of latency and reliability requirements. It is observed that a higher energy consumption is incurred when the UEs have a more strict latency deadline, since the UEs in this case need to use a higher transmission power in order to satisfy their latency requirement. In addition, an increase in the energy consumption is caused by a higher reliability requirement, since the UEs need to offload the tasks to more cloudlets with lower quality channels, which forces the UEs to allocate higher transmission power to meet their latency deadline. Furthermore, it can be seen that when the operator uses a larger IRS, the increase in the UEs' transmission power and energy consumption can be significantly counteracted by selectively improving the channels' quality in response to stricter service requirements.

\begin{figure}[!t]
    \centering
    \includegraphics[width=.5\textwidth]{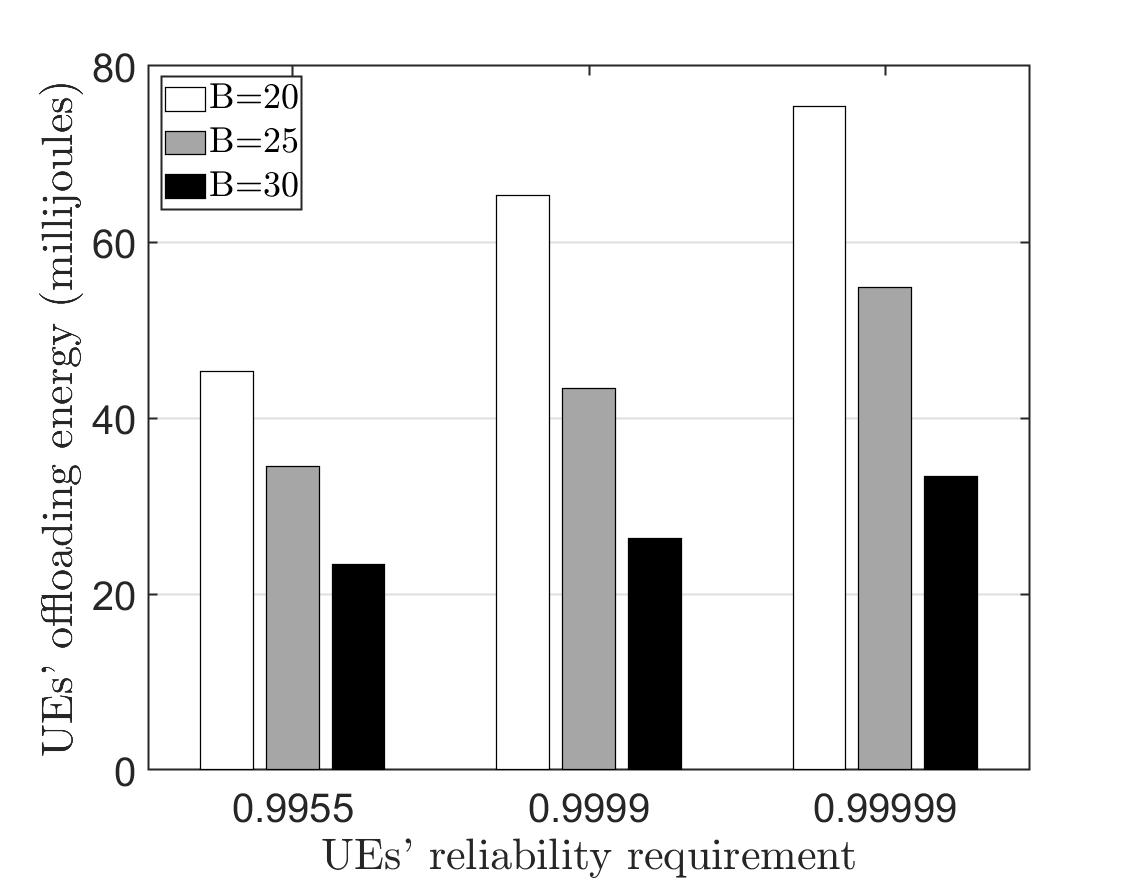}
    \caption{Objective value vs reliability requirement and number of resource blocks}
    \label{fig:plot5RelB}
\end{figure}
\blue{In Fig. \ref{fig:plot5RelB}, we examine the impact of the available system bandwidth on UEs' energy consumption across various reliability levels, utilizing a simpler instance that consists of $K=10$ UEs, $M=3$ randomly distributed APs, and $N=30$ elements for each IRS. This is to showcase the algorithm's performance and adaptability in a more condensed network environment, while providing more convenience due to the scaled-down network model. It captures the nuanced impacts on energy consumption under diverse bandwidth and reliability conditions, without loss of generality.} As can be seen, higher energy is exerted when a more stringent reliability is demanded, and this is because the UEs use an increased transmission power to reach more APs. However, the availability of more RBs can help in maintaining the needed transmission rate, and hence avoids consuming more energy in response to an increased reliability requirement.

\begin{figure}[!t]
    \centering
    \includegraphics[width=.5\textwidth]{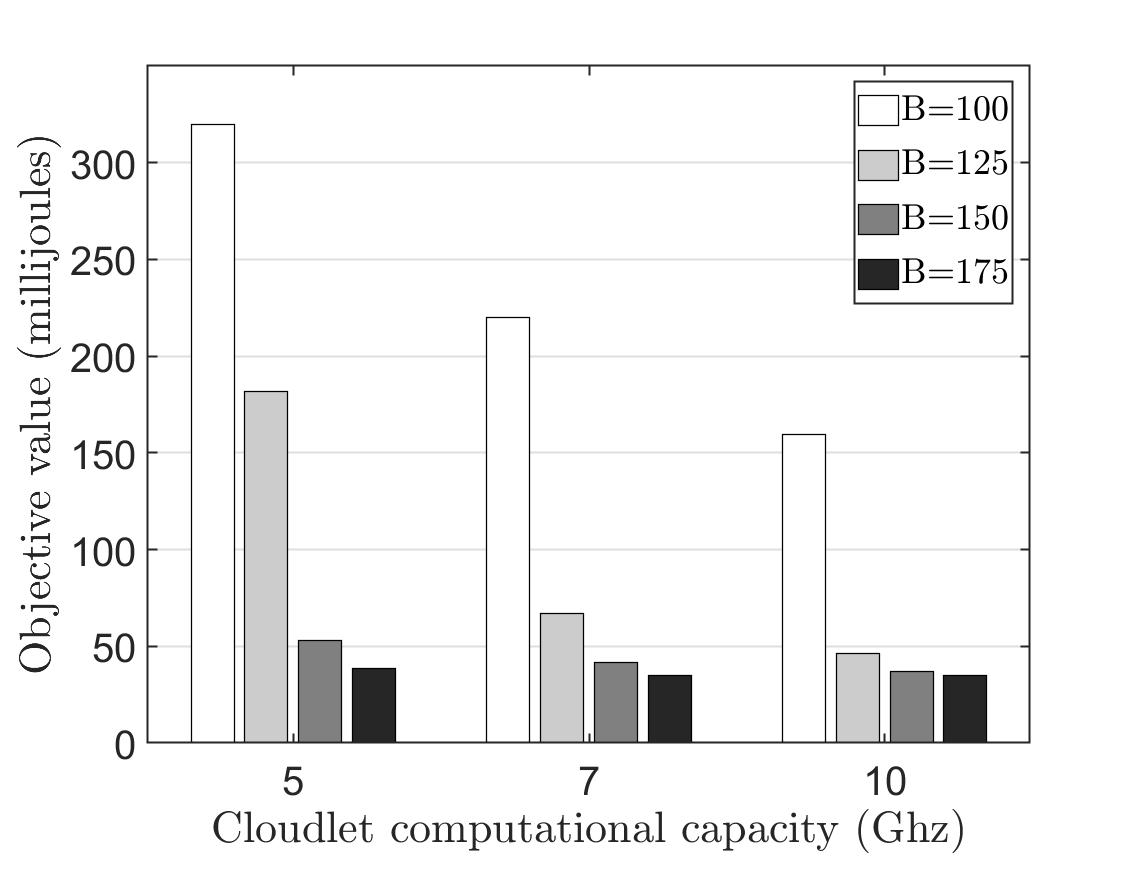}
    \caption{\blue{Objective value vs data size and latency requirement for an instance with heterogeneous tasks.}}
    \label{fig:plot6heteroTasks}
\end{figure}
\blue{In Fig. \ref{fig:plot6heteroTasks}, we examine the performance of our proposed solution in a scenario involving heterogeneous tasks. Here, the objective value is influenced by the availability of network resources, namely the cloudlet computational capacity ($F_j$) and the number of available RBs ($B$). The UEs have tasks that differ significantly in terms of data size ($D_k$) and latency requirements ($L_k$). Specifically, the data sizes ($D_k$) are randomly generated from a uniform distribution in the range [10, 100] kilobits, and the latency requirements ($L_k$) are similarly drawn from a uniform distribution within the range [10, 100] milliseconds. This variation in tasks mirrors real-world scenarios, where offloaded tasks come from different applications with diverse characteristics and requirements. As observed, the UEs can achieve lower energy consumption when more communication and computation resources are available in the network. This energy saving is more pronounced when the network is constrained by resources, and becomes less significant in scenarios where network resources are abundant, indicating a state of saturation in network resources.}

\begin{figure}[!t]
    \centering
    \includegraphics[width=.5\textwidth]{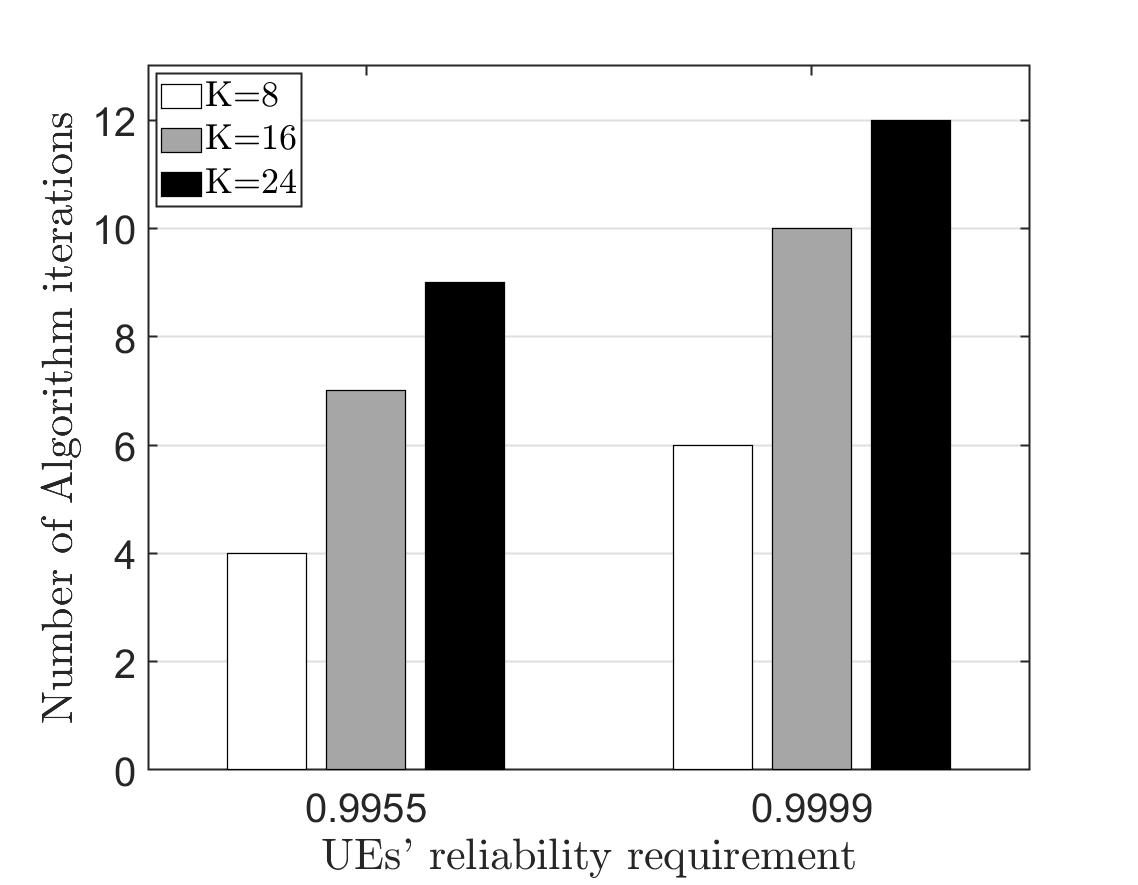}
    \caption{\blue{Algorithm convergence vs number of UEs and reliability requirement.}}
    \label{fig:plot7algIter}
\end{figure}
\blue{In Fig. \ref{fig:plot7algIter}, we illustrate the number of iterations needed for our proposed algorithm to converge when the number of UEs increases, grouped by $2$ levels of UEs' reliability requirement, reflecting the complexity of the simulated instance. As it can be seen, the algorithm is able to converge in a reasonable number of iterations. The convergence needs more iterations whenever the number of UEs or their reliability requirement increases, which reflects the impact of the increased instance's complexity on the convergence of the algorithm. It is worth noting here that the algorithm has a polynomial complexity as shown at the end of Section \ref{sec:solution}, and that the algorithm's runtime is highly dependent on the particular infrastructure and resources used for running it in real-world implementations.}

\section{Conclusion}\label{sec:conclusion}
In this paper, we studied an IRS-assisted MEC system for enabling low-energy computation offloading for IoT services with strict latency and reliability requirements. We presented our problem for optimizing the offloading decision, the IRS elements' phase shifts, and the resources' allocation considering redundancy of task computation for minimizing the UEs' total offloading energy consumption. Being non-convex, the problem is decomposed to separately optimize the problem decisions. First, the offloading decisions are optimized, and then the sub-problems of IRS elements' phase shifts and resources allocation are solved separately in an alternating fashion through algorithms based on the DC and SCA approaches. Through numerical results and owing to the optimized use of IRSs, we demonstrated the savings in energy as well as in network resources when offloading high reliability services, and we highlighted the IRSs' influence on the design of the MEC parameters. Our work offers valuable insights into providing computation offloading support for services with strict latency and reliability requirements with the aid of a deployed IRSs. In our upcoming work, we will study the IRS-aided computation offloading problem considering uncertainty in the incoming requests, prompting us to utilize a machine learning approach to solve the problem.

\begin{appendices}
    \section{Convexification Steps of \eqref{pr:reliabConv}}\label{app:convReliab}
    First, we re-arrange the terms of \eqref{cons:reliab} as:
    \begin{equation}
        \label{pra:reliab_conv1}
        1-\bar{R}_k \geq \prod_{j \in \mathcal{M}}\left(1-x_{kj}\phi_j\right)
    \end{equation}
    Then, after taking the natural logarithm of both sides and using its properties, constraint \eqref{pra:reliab_conv1} can be equivalently rewritten as:
    \begin{equation}
        \label{pra:reliab_conv2}
        \ln(1-\bar{R}_k) \geq \sum_{j \in \mathcal{M}}\ln(1-x_{kj}\phi_j)
    \end{equation}
    The Right-hand Side (RHS) of \eqref{pra:reliab_conv2} $\forall \ j \in \mathcal{M}$ is equal to $\ln(1-\phi_j)$ if $x_{kj}=1$, and to $0$ if $x_{kj}=0$.  Thus, \eqref{pra:reliab_conv2} can be replaced by the following linear constraint:
    \begin{equation}
        \label{pra:reliab_conv}
        \ln(1-\bar{R}_k) \geq \sum_{j \in \mathcal{M}}x_{kj}\ln(1-\phi_j)
    \end{equation}
\end{appendices}

\bibliographystyle{IEEEtran}
\bibliography{IEEEabrv,main}

\begin{thebibliography}{10}
\providecommand{\url}[1]{#1}
\csname url@samestyle\endcsname
\providecommand{\newblock}{\relax}
\providecommand{\bibinfo}[2]{#2}
\providecommand{\BIBentrySTDinterwordspacing}{\spaceskip=0pt\relax}
\providecommand{\BIBentryALTinterwordstretchfactor}{4}
\providecommand{\BIBentryALTinterwordspacing}{\spaceskip=\fontdimen2\font plus
\BIBentryALTinterwordstretchfactor\fontdimen3\font minus \fontdimen4\font\relax}
\providecommand{\BIBforeignlanguage}[2]{{%
\expandafter\ifx\csname l@#1\endcsname\relax
\typeout{** WARNING: IEEEtran.bst: No hyphenation pattern has been}%
\typeout{** loaded for the language `#1'. Using the pattern for}%
\typeout{** the default language instead.}%
\else
\language=\csname l@#1\endcsname
\fi
#2}}
\providecommand{\BIBdecl}{\relax}
\BIBdecl

\bibitem{Elie_ICC}
E.~El~Haber, M.~Elhattab, C.~Assi, S.~Sharafeddine, and K.~K. Nguyen, ``{Latency and Reliability Aware Edge Computation Offloading in IRS-aided Networks},'' in \emph{2022 IEEE International Conference on Communications (ICC)}, Jun. 2022, pp. 1--6.

\bibitem{siddiqi20195g}
M.~A. Siddiqi and al., ``{5G ultra-reliable low-latency communication implementation challenges and operational issues with IoT devices},'' \emph{Electronics}, vol.~8, no.~9, p. 981, 2019.

\bibitem{mao2017survey}
Y.~Mao and al., ``{A survey on mobile edge computing: The communication perspective},'' \emph{IEEE Communications Surveys \& Tutorials}, 2017.

\bibitem{hu2015mobile}
Y.~C. Hu and al., ``{Mobile edge computing—A key technology towards 5G},'' \emph{ETSI white paper}, vol.~11, no.~11, pp. 1--16, 2015.

\bibitem{yazid2021uav}
Y.~Yazid, I.~Ez-Zazi, A.~Guerrero-Gonz{\'a}lez, A.~El~Oualkadi, and M.~Arioua, ``Uav-enabled mobile edge-computing for iot based on ai: A comprehensive review,'' \emph{Drones}, vol.~5, no.~4, p. 148, 2021.

\bibitem{wu2021intelligent}
Q.~Wu and al., ``{Intelligent reflecting surface aided wireless communications: A tutorial},'' \emph{IEEE Transactions on Communications}, 2021.

\bibitem{wu2019towards}
------, ``{Towards smart and reconfigurable environment: Intelligent reflecting surface aided wireless network},'' \emph{IEEE Communications Magazine}, vol.~58, no.~1, pp. 106--112, 2019.

\bibitem{wu2019intelligent}
Q.~Wu, , and al., ``{Intelligent reflecting surface enhanced wireless network via joint active and passive beamforming},'' \emph{IEEE Transactions on Wireless Communications}, vol.~18, no.~11, pp. 5394--5409, 2019.

\bibitem{marks1978general}
B.~R. Marks and al., ``A general inner approximation algorithm for nonconvex mathematical programs,'' \emph{Operations research}, vol.~26, no.~4, pp. 681--683, 1978.

\bibitem{10109654}
S.~Khisa, M.~Elhattab, C.~Assi, and S.~Sharafeddine, ``Energy consumption optimization in ris-assisted cooperative rsma cellular networks,'' \emph{IEEE Transactions on Communications}, vol.~71, no.~7, pp. 4300--4312, July 2023.

\bibitem{9586734}
M.~Elhattab, M.~A. Arfaoui, C.~Assi, and A.~Ghrayeb, ``Reconfigurable intelligent surface enabled full-duplex/half-duplex cooperative non-orthogonal multiple access,'' \emph{IEEE Transactions on Wireless Communications}, vol.~21, no.~5, pp. 3349--3364, May 2022.

\bibitem{bai2020latency}
T.~Bai and al., ``{Latency minimization for intelligent reflecting surface aided mobile edge computing},'' \emph{IEEE Journal on Selected Areas in Communications}, vol.~38, no.~11, pp. 2666--2682, 2020.

\bibitem{liu2020intelligent}
Y.~Liu and al., ``{Intelligent reflecting surface meets mobile edge computing: Enhancing wireless communications for computation offloading},'' \emph{arXiv preprint arXiv:2001.07449}, 2020.

\bibitem{bai2021resource}
T.~Bai and al., ``{Resource allocation for intelligent reflecting surface aided wireless powered mobile edge computing in {OFDM} systems},'' \emph{IEEE Transactions on Wireless Communications}, pp. 1--1, Mar 2021.

\bibitem{chu2020intelligent}
Z.~Chu and al., ``{Intelligent Reflecting Surface Assisted Mobile Edge Computing for Internet of Things},'' \emph{IEEE Wireless Communications Letters}, 2020.

\bibitem{li2021energy}
Z.~Li and al., ``{Energy Efficient Reconfigurable Intelligent Surface Enabled Mobile Edge Computing Networks with {NOMA}},'' \emph{IEEE Transactions on Cognitive Communications and Networking}, Mar 2021.

\bibitem{wu2021irs}
Q.~Wu, W.~Chen, D.~W.~K. Ng, L.~Hanzo \emph{et~al.}, ``{IRS-aided Wireless Powered MEC Systems: TDMA or NOMA for Computation Offloading?}'' \emph{arXiv preprint arXiv:2108.06120}, 2021.

\bibitem{liu2017latency}
C.-F. Liu and al., ``{Latency and reliability-aware task offloading and resource allocation for mobile edge computing},'' in \emph{Globecom Workshops (GC Wkshps), 2017 IEEE}.\hskip 1em plus 0.5em minus 0.4em\relax IEEE, 2017, pp. 1--7.

\bibitem{liu2018offloading}
J.~Liu and al., ``{Offloading schemes in mobile edge computing for ultra-reliable low latency communications},'' \emph{IEEE Access}, vol.~6, pp. 12\,825--12\,837, 2018.

\bibitem{xu2022energy}
Z.~Xu, J.~Liu, J.~Zou, and Z.~Wen, ``Energy-efficient design for irs-assisted noma-based mobile edge computing,'' \emph{IEEE Communications Letters}, 2022.

\bibitem{zhou2022latency}
Y.~Zhou, C.~Pan, P.~L. Yeoh, K.~Wang, Z.~Ma, B.~Vucetic, and Y.~Li, ``Latency minimization for secure intelligent reflecting surface enhanced virtual reality delivery systems,'' \emph{IEEE Wireless Communications Letters}, 2022.

\bibitem{hu2021reconfigurable}
X.~Hu, C.~Masouros, and K.-K. Wong, ``Reconfigurable intelligent surface aided mobile edge computing: From optimization-based to location-only learning-based solutions,'' \emph{IEEE Transactions on Communications}, vol.~69, no.~6, pp. 3709--3725, 2021.

\bibitem{yang2022intelligent}
Y.~Yang, Y.~Gong, and Y.-C. Wu, ``Intelligent reflecting surface aided mobile edge computing with binary offloading: Energy minimization for iot devices,'' \emph{IEEE Internet of Things Journal}, 2022.

\bibitem{mao2022reconfigurable}
S.~Mao, L.~Liu, N.~Zhang, M.~Dong, J.~Zhao, J.~Wu, and V.~C. Leung, ``Reconfigurable intelligent surface-assisted secure mobile edge computing networks,'' \emph{IEEE Transactions on Vehicular Technology}, 2022.

\bibitem{zhang2021drl}
X.~Zhang, Y.~Shen, B.~Yang, W.~Zang, and S.~Wang, ``Drl based data offloading for intelligent reflecting surface aided mobile edge computing,'' in \emph{2021 IEEE Wireless Communications and Networking Conference (WCNC)}.\hskip 1em plus 0.5em minus 0.4em\relax IEEE, 2021, pp. 1--7.

\bibitem{9563047}
E.~El~Haber, M.~Elhattab, C.~Assi, S.~Sharafeddine, and K.~K. Nguyen, ``{Latency and Reliability Aware Edge Computation Offloading via an Intelligent Reflecting Surface},'' \emph{IEEE Communications Letters}, pp. 1--1, 2021.

\bibitem{wang2017joint}
C.~Wang and al., ``{Joint computation offloading and interference management in wireless cellular networks with mobile edge computing},'' \emph{IEEE Transactions on Vehicular Technology}, vol.~66, no.~8, pp. 7432--7445, 2017.

\bibitem{Wei_2021_Channel}
\color{black} L.~Wei~\textit{et al.}, ``Channel estimation for {RIS}-empowered multi-user {MISO} wireless communications,'' \emph{{IEEE} Trans. Commun.}, vol.~69, no.~6, pp. 4144--4157, Jun. 2021.

\bibitem{taha2019enabling}
A.~Taha~\textit{et al.}, ``{Enabling Large Intelligent Surfaces With Compressive Sensing and Deep Learning},'' \emph{IEEE Access}, vol.~9, pp. 44\,304 -- 44\,321, Mar. 2021.

\bibitem{9388935}
T.~Bai, C.~Pan, H.~Ren, Y.~Deng, M.~Elkashlan, and A.~Nallanathan, ``Resource allocation for intelligent reflecting surface aided wireless powered mobile edge computing in ofdm systems,'' \emph{IEEE Transactions on Wireless Communications}, vol.~20, no.~8, pp. 5389--5407, 2021.

\bibitem{kherraf2019latency}
N.~Kherraf and al., ``{Latency and reliability-aware workload assignment in {IoT} networks with mobile edge clouds},'' \emph{IEEE Transactions on Network and Service Management}, pp. 1435--1449, Oct 2019.

\bibitem{el2021uav}
E.~El~Haber, H.~A. Alameddine, C.~Assi, and S.~Sharafeddine, ``Uav-aided ultra-reliable low-latency computation offloading in future iot networks,'' \emph{IEEE Transactions on Communications}, vol.~69, no.~10, pp. 6838--6851, 2021.

\bibitem{vishwanath2010characterizing}
K.~V. Vishwanath and al., ``{Characterizing cloud computing hardware reliability},'' in \emph{Proceedings of the 1st ACM symposium on Cloud computing}, Jun 2010, pp. 193--204.

\bibitem{zhang2014venice}
Q.~Zhang and al., ``Venice: Reliable virtual data center embedding in clouds,'' in \emph{INFOCOM, 2014 Proceedings IEEE}.\hskip 1em plus 0.5em minus 0.4em\relax IEEE, 2014, pp. 289--297.

\bibitem{ayoubi2016reliable}
S.~Ayoubi, Y.~Zhang, and C.~Assi, ``A reliable embedding framework for elastic virtualized services in the cloud,'' \emph{IEEE Transactions on Network and Service Management}, vol.~13, no.~3, pp. 489--503, 2016.

\bibitem{kazmi2017mode}
S.~A. Kazmi, N.~H. Tran, W.~Saad, Z.~Han, T.~M. Ho, T.~Z. Oo, and C.~S. Hong, ``{Mode selection and resource allocation in device-to-device communications: A matching game approach},'' \emph{IEEE Transactions on Mobile Computing}, vol.~16, no.~11, pp. 3126--3141, 2017.

\bibitem{ben2001polyhedral}
A.~Ben-Tal and al., ``{On polyhedral approximations of the second-order cone},'' \emph{Mathematics of Operations Research}, vol.~26, no.~2, pp. 193--205, 2001.

\bibitem{so2007approximating}
A.~M.-C. So and al., ``{On approximating complex quadratic optimization problems via semidefinite programming relaxations},'' \emph{Mathematical Programming}, vol. 110, no.~1, pp. 93--110, 2007.

\bibitem{10225434}
A.~Muhammad, M.~Elhattab, M.~A. Arfaoui, and C.~Assi, ``Optimizing age of information in ris-empowered uplink cooperative noma networks,'' \emph{IEEE Transactions on Network and Service Management}, pp. 1--1, 2023.

\bibitem{nguyen2018novel}
T.~M. Nguyen, W.~Ajib, and C.~Assi, ``A novel cooperative non-orthogonal multiple access ({NOMA}) in wireless backhaul two-tier hetnets,'' \emph{IEEE Transactions on Wireless Communications}, 2018.

\bibitem{polik2010interior}
I.~P{\'o}lik and T.~Terlaky, ``{Interior point methods for nonlinear optimization},'' in \emph{Nonlinear optimization}.\hskip 1em plus 0.5em minus 0.4em\relax Springer, Jan 2010, pp. 215--276.

\end{thebibliography}

\end{document}